\documentclass[fleqn,usenatbib]{mnras}

\usepackage{newtxtext,newtxmath}

\usepackage[T1]{fontenc}

\DeclareRobustCommand{\VAN}[3]{#2}
\let\VANthebibliography\thebibliography
\def\thebibliography{\DeclareRobustCommand{\VAN}[3]{##3}\VANthebibliography}

\usepackage{booktabs}
\usepackage{graphicx}	
\usepackage{amsmath}	
\usepackage{layouts}
\usepackage{gensymb}
\usepackage{hyperref}
\usepackage{bm}
\usepackage[encapsulated]{CJK}

\newcommand{\qF}{\textit{m10q}}
\newcommand{\qA}{\textit{m10qAGN}}
\newcommand{\qC}{\textit{m10qCR}}
\newcommand{\qCA}{\textit{m10qCR+AGN}}
\newcommand{\yF}{\textit{m10y}}
\newcommand{\yA}{\textit{m10yAGN}}
\newcommand{\yC}{\textit{m10yCR}}
\newcommand{\yCA}{\textit{m10yCR+AGN}}
\newcommand{\zF}{\textit{m10z}}
\newcommand{\zA}{\textit{m10zAGN}}
\newcommand{\zC}{\textit{m10zCR}}
\newcommand{\zCA}{\textit{m10zCR+AGN}}

\newcommand{\Msun}{\mathrm{M_{\odot}}}

\newcommand{\pkg}[1]{\texttt{#1}}

\newcommand{\Zsun}{\mathrm{Z}_\mathrm{\odot}}

\newcommand{\Mgas}{M_\mathrm{gas}}
\newcommand{\Mvir}{M_\mathrm{vir}}

\newcommand{\Mdotbh}{\dot{M}_\mathrm{BH}}

\newcommand{\Mdotacc}{\dot{M}_\mathrm{acc}}

\newcommand{\vwind}{v_\mathrm{wind}}

\newcommand{\kms}{\mathrm{km}\,\mathrm{s}^{-1}}

\newcommand{\Mdm}{M_\mathrm{DM}}
\newcommand{\Mbh}{M_\mathrm{BH}}

\title[Cusp versus core with AGN and cosmic rays]{Diverse dark matter profiles in \textsc{fire} dwarfs: black holes, cosmic rays and the cusp--core enigma}

\author[S. Koudmani et al.]{Sophie Koudmani,$^{1,2,3,4}$\thanks{E-mail: skoudmani@ast.cam.ac.uk}
Douglas Rennehan,$^{3}$
Rachel S. Somerville,$^{3}$
Christopher C. Hayward,$^{3}$
\newauthor
Daniel Anglés-Alcázar,$^{5,3}$
Matthew E. Orr,$^{3}$
Isabel S. Sands$^{6}$
and Sarah Wellons$^{7}$
\\
$^{1}$ St Catharine's College, University of Cambridge, Trumpington Street, Cambridge CB2 1RL, UK\\
$^{2}$ Institute of Astronomy and Kavli Institute for Cosmology, University of Cambridge, Madingley Road, Cambridge, CB3 0HA, UK\\
$^{3}$ Center for Computational Astrophysics, Flatiron Institute, 162 Fifth Avenue, New York, NY 10010, USA\\
$^{4}$ Centre for Astrophysics Research, Department of Physics, Astronomy and Mathematics, University of Hertfordshire, College Lane, Hatfield, AL10 9AB, UK\\
$^{5}$ Department of Physics, University of Connecticut, 196 Auditorium Road, U-3046, Storrs, CT 06269, USA \\
$^{6}$ TAPIR, California Institute of Technology, Mailcode 350-17, Pasadena, CA 91125, USA \\
$^{7}$ Department of Astronomy, Van Vleck Observatory, Wesleyan University, 96 Foss Hill Drive, Middletown, CT 06459, USA\\
}

\date{Submitted to MNRAS}

\pubyear{2024}

\begin{document}
\label{firstpage}
\pagerange{\pageref{firstpage}--\pageref{lastpage}}
\maketitle

\begin{abstract}
Dwarf galaxies have historically posed challenges to the cold dark matter (CDM) model and, while many of the so-called `dwarf galaxy problems' have been mitigated by incorporating baryonic processes, the observed diversity of dwarf galaxy rotation curves remains a contentious topic. Meanwhile, the growing observational samples of active galactic nuclei (AGN) in dwarf galaxies have prompted a paradigm shift in our understanding of dwarf galaxy evolution, traditionally thought to be regulated by stellar feedback. In this study, we explore the potential role of AGN feedback in shaping dark matter distributions and increasing the diversity of dwarf galaxy rotation curves, using a new suite of cosmological zoom-in simulations of dwarf galaxies with the \textsc{fire-3} model. Our findings indicate that the presence of active black holes (BHs) in dwarf galaxies can lead to diverse outcomes, ranging from cuspier to more core-like profiles. This variability arises from the dual role of BHs in providing additional feedback and regulating the extent of stellar feedback. Consistent with previous research, we find that AGN feedback is most impactful when cosmic ray (CR) modelling is included, with CRs from any source significantly influencing dark matter profiles. Overall, our results highlight that the interplay between stellar feedback, BHs, and CRs produces a broad spectrum of dark matter density profiles, which align with observed correlations between rotation curve shapes and baryonic dominance. This underscores the importance of including the full range of baryonic processes in dwarf galaxy simulations to address the persistent `small-scale challenges' to the CDM paradigm.
\end{abstract}




\section{Introduction} \label{sec:intro}

Dwarf galaxies are pivotal in the Lambda -- Cold Dark Matter ($\Lambda$CDM) model, serving as the building blocks of hierarchical structure formation. Defined as galaxies with stellar masses below $3 \times 10^{9} \ \Msun$ (similar to the mass of the Large Magellanic Cloud), dwarf galaxies are crucial astrophysical probes in near-field cosmology and galaxy formation studies.

Due to their shallow potential wells, dwarf galaxies are ideal testbeds for studying galactic feedback processes (`baryonic physics'). They are also highly dark-matter dominated, making them valuable for testing dark matter models. However, their sensitivity to both dark matter and baryonic processes has led to several controversies, collectively known as the `dwarf galaxy problems', which highlight discrepancies between observations and $\Lambda$CDM predictions from N-body simulations \citep[see][for a recent review]{sales_baryonic_2022}. Key issues include the missing satellites problem, which notes a deficit in the observed number of Milky Way satellites compared to theoretical predictions \citep[e.g.][]{kauffmann_formation_1993,klypin_where_1999,moore_dark_1999}, and the too-big-to-fail problem, where the central masses of the largest observed satellites do not match those of the most massive simulated subhaloes \citep[e.g.][]{boylan-kolchin_too_2011}. Additionally, the cusp-core problem refers to the observation that dwarf galaxy dark matter halo profiles are less centrally peaked or `cuspy' than predicted by the CDM model with some dwarf galaxies instead exhibiting core-like profiles \citep[e.g.][]{flores_observational_1994,moore_evidence_1994}.

These discrepancies have led to explorations of alternative dark matter models, such as warm dark matter (WDM) \citep{blumenthal_galaxy_1982}. WDM models, which include a free-streaming cut-off at dwarf galaxy scales, can address the missing satellite problem and reduce central halo densities, thus partially resolving the too-big-to-fail problem \citep{lovell_haloes_2012}. However, WDM struggles with the cusp-core problem, as the resulting cores are not large enough \citep[e.g.][]{shao_phase-space_2013}. Another approach involves self-interacting dark matter (SIDM), where particles scatter elastically, creating cored density profiles and reducing halo ellipticity \citep[see][for a review]{tulin_dark_2018}. Constraints from strong lensing measurements of galaxy clusters suggest a velocity-dependent cross section that decreases from dwarf to cluster scales \citep[e.g.][]{meneghetti_giant_2001, randall_constraints_2008, vogelsberger_subhaloes_2012}. Additionally, fuzzy dark matter (FDM), which proposes extremely low-mass particles like ultralight axions, produces cored central profiles and suppresses small-scale structures in the dwarf regime due to their long de Broglie wavelengths \citep[see][for a review]{niemeyer_small-scale_2020}.

However, there has also been a large body of theoretical work focusing on improving the baryonic physics modelling which may largely alleviate the `dwarf galaxy problems' observed in pure CDM simulations. In particular, star formation suppression from reionization could significantly decrease the number of bright dwarf satellites \citep[e.g.][]{efstathiou_suppressing_1992,okamoto_mass_2008,fitts_fire_2017,katz_how_2020} thereby resolving the missing satellite problem. Furthermore, stellar feedback \citep[e.g.][]{navarro_cores_1996,governato_bulgeless_2010,parry_baryons_2012,brooks_why_2014,hopkins_galaxies_2014,hopkins_fire-2_2018,chan_impact_2015,kimm_towards_2015,emerick_stellar_2018,smith_supernova_2018,smith_cosmological_2019,gutcke_lyra_2021,jackson_formation_2024} may significantly decrease the central dark matter densities of dwarf galaxies. In particular, it has been demonstrated that cyclic supernova (SN) bursts play an important role in creating cores in dwarf galaxies \citep[e.g.][]{pontzen_how_2012} and that late-time star formation is crucial for maintaining these cores \citep[e.g.][]{read_dark_2019,muni_dark_2024}. 

Notwithstanding this progress in alternative dark matter models and improved baryonic physics, some prominent discrepancies between simulations and observations still appear difficult to resolve without fine-tuning the models. In particular, theoretical models significantly struggle with explaining the observed diversity of dwarf rotation curves \citep[see e.g.][for a recent compilation]{santos-santos_baryonic_2020}. Due to the self-similar nature of structure formation in the $\Lambda$CDM universe, the full mass profile of a halo can be characterised by a single parameter such as the virial mass \citep{navarro_universal_1997}. Hence dwarf galaxy rotation curve shapes are predicted to be almost identical for similar maximum circular velocities -- and yet a great diversity is observed ranging from extremely cuspy profiles to large cores for the same inferred dark matter halo mass \citep[see][]{oman_unexpected_2015}. Galaxy formation simulations based on CDM have traditionally struggled to reproduce this observed diversity of dwarf rotation curves \citep{dutton_nihao_2016-1,sawala_apostle_2016,garrison-kimmel_local_2019,applebaum_ultrafaint_2021}. Some groups find that SIDM may be able to improve these discrepancies \citep[e.g.][]{ren_reconciling_2019,kaplinghat_dark_2020}, whilst others find no preference between SIDM and CDM simulations with stellar feedback \citep[e.g.][]{santos-santos_baryonic_2020,zentner_critical_2022}.

The most fundamental issue for CDM simulations lies in obtaining a balance in the burstiness of star formation histories, with some simulation set-ups predominantly producing smooth (or quenched) star formation histories and cusps \citep[e.g.][]{vogelsberger_introducing_2014,sawala_apostle_2016,gutcke_lyra_2022} whilst others predominantly produce bursty star formation histories and cores \citep[e.g.][]{navarro_cores_1996,read_mass_2005,governato_bulgeless_2010,pontzen_how_2012,chan_impact_2015,garrison-kimmel_local_2019}. 

One potential solution could be additional feedback processes in the dwarf galaxy regime interacting with SNe and thereby modulating the star formation histories, leading to a natural diversity in rotation curves \citep[see discussion in][]{garrison-kimmel_local_2019}. Recently, the role of active galactic nuclei (AGN) in resolving the remaining dwarf galaxy problems has garnered significant interest. Traditionally, AGN feedback was only considered in massive galaxies, with stellar feedback thought to dominate in low-mass galaxies. However, these models have been called into question by the growing observational samples of AGN in dwarfs, with detections spanning the whole electromagnetic spectrum from X-ray \citep[e.g.][]{schramm_unveiling_2013,baldassare_50000_2015,baldassare_x-ray_2017,lemons_x-ray_2015,miller_x-ray_2015,mezcua_population_2016,mezcua_intermediate-mass_2018,pardo_x-ray_2016,aird_x-rays_2018,birchall_x-ray_2020,birchall_incidence_2022,latimer_agn_2021} to optical \citep[e.g.][]{greene_active_2004,greene_new_2007,reines_dwarf_2013,chilingarian_population_2018,molina_sample_2021,polimera_resolve_2022} to IR \citep[e.g.][]{satyapal_discovery_2014,sartori_search_2015,marleau_infrared_2017,kaviraj_agn_2019} and to radio observations \citep[e.g.][]{greene_radio_2006,wrobel_radio_2006,wrobel_steep-spectrum_2008,nyland_intermediate-mass_2012,nyland_multi-wavelength_2017,reines_parsec-scale_2012,reines_candidate_2014,mezcua_extended_2018,mezcua_radio_2019,reines_new_2020,davis_radio_2022}. Intriguingly, recent targeted surveys indicate high AGN occupation fractions, on the order of 10 per cent, in dwarf galaxies, suggesting there may not be a drastic decline of AGN activity in the dwarf regime \citep[e.g.][]{baldassare_50000_2015,birchall_incidence_2022,bichanga_properties_2024,mezcua_manga_2024}.

These tantalising observations have motivated theorists to investigate this largely unexplored regime. Analytical models point towards favourable AGN energetics in the dwarf regime \citep[compared to stellar feedback, see e.g.][]{dashyan_agn_2018} and suggest that AGN activity may be able resolve all of the remaining `dwarf galaxy problems' \citep{silk_feedback_2017}. However, hydrodynamical simulations have painted a more complex picture. AGN activity in dwarf galaxies may be significantly suppressed by stellar feedback evacuating gas from the central region \citep[e.g.][]{dubois_black_2015,angles-alcazar_black_2017,habouzit_blossoms_2017,trebitsch_escape_2018,hopkins_why_2022,byrne_stellar_2023} and black holes (BHs) may be wandering in dwarf galaxies due to their shallow potential wells \citep[e.g.][]{bellovary_origins_2021,ma_seeds_2021,sharma_hidden_2022,beckmann_population_2023} further decreasing their accretion efficiency. The modelling of BH growth is another crucial aspect as the fiducial Bondi model suppresses the growth of low-mass BHs compared to gas-supply-limited or torque-driven schemes due to its strong dependence on BH mass \citep[e.g.][]{angles-alcazar_black_2013,angles-alcazar_torque-limited_2015,koudmani_two_2022,wellons_exploring_2023,gordon_hungry_2024}. The cosmic environment also has a crucial influence on AGN activity in the dwarf regime with minor mergers triggering AGN episodes whilst long-term residence in dense environments is detrimental to BH growth \citep[e.g.][]{kristensen_merger_2021}.

Despite these caveats, various groups have identified physical regimes where AGN in dwarfs can accrete efficiently in accordance with observed samples. These dwarf AGN significantly influence their host galaxies by driving powerful outflows \citep[e.g.][]{koudmani_fast_2019,koudmani_two_2022} and suppressing star formation \citep[e.g.][]{barai_intermediate-mass_2019,sharma_black_2020,sharma_hidden_2022,sharma_active_2023,koudmani_little_2021,koudmani_two_2022,wellons_exploring_2023,arjona-galvez_role_2024}. Given this potentially significant impact by AGN on the baryon cycle in dwarf galaxies, could AGN feedback also impact the dark matter distributions in dwarfs as predicted by analytical models? 

The interaction between AGN feedback, star formation, and dark matter in dwarf galaxies has not been thoroughly explored, though some trends have been noted, such as AGN mildly suppressing central dark matter densities \citep[see][]{koudmani_two_2022,arjona-galvez_role_2024}, hinting at a possible role of AGN in driving cusp-to-core transformations in dwarfs. However, the central density suppressions by the AGN in these simulations were mostly minor and did not change the qualitative nature of the central profiles. This is likely due to the effective equation of state \citep{springel_cosmological_2003} employed for modelling the interstellar medium (ISM) in these simulations, which leads to smoother star formation histories alongside less bursty AGN feedback at low redshifts, often operating as `maintenance-mode' feedback \citep[also see discussion in][]{koudmani_two_2022}. Recently, \citet{arora_dark_2024} examined the role of BH feedback in core formation within the NIHAO \citep{wang_nihao_2015} simulations and found negligible effects in low-mass systems, largely due to inefficient BH growth in this mass regime, constrained by the Bondi accretion scheme. In contrast, they found significant impacts of BH feedback on core formation in more massive galaxies, aligning with earlier studies that show efficient AGN activity can create cores in these galaxies \citep[e.g.][]{martizzi_cusp-core_2013,peirani_density_2017,maccio_nihao_2020}.

It is therefore timely to investigate whether efficient AGN activity can induce cusp-to-core transformations in dwarf galaxies and potentially account for the observed diversity in dwarf rotation curves. Clearly, within this context it is crucial to model the multi-phase nature of the ISM but also to consider the role of non-thermal components such as magnetic fields or cosmic rays (CRs) \citep[e.g.][]{uhlig_galactic_2012,booth_simulations_2013,hanasz_cosmic_2013,salem_cosmic_2014,pakmor_galactic_2016,farber_impact_2018,holguin_role_2019,dashyan_cosmic_2020,martin-alvarez_how_2020,steinwandel_origin_2020}. In particular, it has been shown that CRs may drive core formation \citep[e.g.][]{hopkins_but_2020,martin-alvarez_pandora_2023} though their impact in the dwarf regime remains controversial as gas-rich mergers may re-establish cores in these systems \citep[see discussion in][]{martin-alvarez_pandora_2023}. What is more, it has been found that CRs may significantly boost the efficiency of AGN feedback \citep[e.g.][]{su_which_2021,su_unravelling_2024,byrne_effects_2023,wellons_exploring_2023} by suppressing cold gas accretion more effectively than thermal heating or momentum injection. Therefore, it is important to also assess the role of CRs when investigating cusp-core transformations with AGN.

The aim of this paper is to systematically investigate the impact of AGN feedback and CRs on dark matter profiles in dwarf galaxies and to assess whether incorporating these additional baryonic processes may be the key to resolving the diversity of rotation curve problem. To this end, we perform simulations with and without CRs and/or AGN of three different dwarf haloes with explicit multi-phase ISM physics within the \textsc{fire-3} galaxy formation framework and contrast the resulting dark matter profiles, their cosmic evolution and their connection with different feedback processes.

The remainder of this paper is structured as follows. In Section~\ref{sec:methods}, we introduce our three dwarf simulation set-ups and describe the different variations of the \textsc{fire-3} galaxy formation model explored in this work. The results of our analysis are presented in Section~\ref{sec:results}. Firstly, we focus on the stellar and BH properties of our simulated dwarfs, including visualisations of the stellar distributions in Section~\ref{subsec:res_overview}, comparisons with observed nearby low-mass galaxies (see Section~\ref{subsec:res_stellar}) and observed BH masses in dwarfs (see Section~\ref{subsec:res_bhs}). We then analyse how our different feedback set-ups influence the dark matter profiles at $z=0$ in Section~\ref{subsec:res_dm} and examine the driving forces behind the diversity of dark matter profiles in \ref{subsec:res_dmforces}. We investigate the cosmic evolution of the central dark matter densities in Section~\ref{subsec:results_centraldm} and assess whether this could explain the observed diversity of rotation curves in Section~\ref{subsec:res_diversity}. We discuss our results and caveats to our modelling in Section~\ref{sec:discussion} and conclude in Section~\ref{sec:conclusions}.

\section{Methodology} \label{sec:methods}

\subsection{General set-up}

Our simulation suite is based on a set of $3$ initial conditions (ICs) of dwarf haloes selected from the \textsc{fire} ICs set, spanning a range of halo masses in the `classical dwarf' regime.  We select two low-mass dwarfs: \textbf{m10q} and \textbf{m10y}, with $z=0$ halo masses of $\Mvir=8\times10^9$ $\Msun$ and $\Mvir=1.4\times10^{10}$ $\Msun$, respectively.  Our third halo, \textbf{m10z}, is an intermediate-mass dwarf with $z=0$ halo mass $\Mvir = 3.5\times10^{10}$ $\Msun$. All of these ICs were generated using the \textsc{music} IC generator \citep{hahn_multi-scale_2011} and have been thoroughly explored with the \textsc{fire-2} model \citep{hopkins_fire-2_2018}. In the standard \textsc{fire-2} runs, \textbf{m10q} has a quiescent growth history and is relatively isolated, \textbf{m10y} forms early and has a large core, and \textbf{m10z} is an ultra-diffuse galaxy. All three of our selected galaxies are central galaxies.

To simulate these systems, we use the \textsc{gizmo} hydrodynamics+gravity code, employing the mesh-free finite mass method \citep{lanson_renormalized_2008,lanson_renormalized_2008-1,gaburov_astrophysical_2011,hopkins_new_2015,hopkins_new_2017} that tracks constant-mass gas resolution elements in a Lagrangian fashion.  Our simulations include both baryons and dark matter, with mass resolutions of $\Mgas = 262$ $\Msun$ and $\Mdm = 1303$ $\Msun$, respectively. The gravitational softening for the dark matter, stars and BH particles is set to $\epsilon_\mathrm{soft} = 0.0625 \ \mathrm{ckpc} \, h^{-1}$. Note that we keep the softening for these particles comoving throughout the simulation. The softening for the gas resolution elements is set adaptively. This allows us to easily resolve core formation as demonstrated in previous work \citep[e.g.][]{chan_impact_2015,hopkins_fire-2_2018}.

\subsection{Galaxy formation model}

For the galaxy formation physics, we use the newly updated \textsc{fire-3} model \citep{hopkins_fire-3_2023} which tracks the state of the multi-phase ISM and multiple forms of stellar feedback, including feedback from SNe\footnote{We couple the SN momentum following \citet{hopkins_fire-3_2023} and do not yet include the updates to the momentum coupling from \citet{hopkins_importance_2024}. Though we note that these changes mainly affect high-mass galaxies at low resolution, whilst the impact on high-resolution dwarfs is very limited.} of Type Ia and II, stellar winds from OB and AGB stars as well as multi-wavelength photo-heating and radiation pressure. Star formation occurs in gas that is molecular, dense, self-gravitating, self-shielding, and Jeans-unstable. All simulations include magnetic fields, using the magneto-hydrodynamics methods from \citet{hopkins_accurate_2016} and \citet{hopkins_constrained-gradient_2016}. The \textsc{fire-3} model also has new optional physics modules compared to the previous \textsc{fire} \citep{hopkins_galaxies_2014} and \textsc{fire-2} \citep{hopkins_fire-2_2018} models. Two of the most important optional features are a novel sub-grid BH model, including seeding, accretion and feedback, as well as the inclusion of CR physics. To separate the impacts of CRs and BHs, we perform four simulations for each set of ICs, including runs without CRs or BHs, with both CRs and BHs and with only CRs or BHs, respectively. For the runs without BHs but with CRs, no BHs are seeded and the CRs only stem from SNe. For the runs without CRs and with BHs, we incorporate all AGN feedback channels except for CR injection and SNe do not inject CRs either. See Table~\ref{tab:dwarfruns} for an overview of the simulation suite. In this table, we also indicate the central dark matter densities of each simulation run measured at $r=150$~pc, a characteristic radius for distinguishing cusps from cores in observations \citep[see][]{read_dark_2019}. For reference, we further provide the ratio between the dark matter densities from the full MHD runs and the dark matter only runs and a qualitative assessment of the central dark matter density shape as weak/strong cusp or core. The latter categorisation is decided by comparing the central densities to the expectations from the fiducial NFW profile and a cored profile following the \textsc{coreNFW} profile from \citet{read_dark_2016}. Weak cusps fall below the mean relation for an NFW profile but within the $2\sigma$ scatter. Similarly, weak cores lie above the \textsc{coreNFW} expectations but within the $2\sigma$ scatter (see Figure~\ref{fig:dm_central_dens}).

\begin{table*}
\caption{Overview of dwarf zoom-in simulation suite. We list the simulation names (first column), ICs (second column), and whether BHs (third column) and CRs (fourth column) are included. We also provide the $z=0$ BH mass (where applicable, see fifth column), central dark matter densities at $r=150$~pc (sixth column), the ratios between the central dark matter density for the hydro runs and dark matter only simulations (seventh column), and the shape of the central dark matter density profile (eighth column) as shown in Figure~\ref{fig:dm_central_dens}. All masses and density estimators are rounded to two significant figures.}
\begin{center}
\begin{tabular}{@{}lccccccc@{}}
\toprule
\textbf{Simulation name} & \textbf{ICs} & \textbf{BHs?} & \textbf{CRs?}  & $\boldsymbol{M_\mathrm{BH} \ [10^{4} \, \mathrm{M_{\odot}}]}$ &  $\boldsymbol{\rho_\mathrm{DM,inner} \ [10^{8} \ \Msun \, \mathrm{kpc^{-3}}]}$ & $\boldsymbol{\rho_\mathrm{DM,inner} / \rho_\mathrm{DMO,inner} }$ & \textbf{Central dark matter profile} \\  \toprule
\textit{m10q}	& m10q	& No	& No & / & 0.73 & 0.47 & weak core \\
\textit{m10qAGN}	& m10q	&Yes	& No	& 2.7 & 0.42&0.27 & strong core \\
\textit{m10qCR} & m10q & No & Yes & / & 1.4& 0.92 & weak cusp	\\		
\textit{m10qCR+AGN} & m10q	& Yes	& Yes	& 0.5 & 1.5 & 0.99& weak cusp \\
\textit{m10y}	& m10y	& No	& No & / & 0.20 & 0.062& strong core \\
\textit{m10yAGN}	& m10y	&Yes	& No & 9.3 & 0.19 & 0.058& strong core \\	
\textit{m10yCR} & m10y & No & Yes & / &0.83 & 0.26& weak core	\\		
\textit{m10yCR+AGN} & m10y	& Yes	& Yes & 2.2	&0.74 &0.23 & weak core \\
\textit{m10z}	& m10z	& No	& No & / &0.22 & 0.10&strong core \\
\textit{m10zAGN}	& m10z	&Yes	& No & 7.0 & 0.069& 0.032& strong core \\
\textit{m10zCR} & m10z & No & Yes & / & 0.11&0.053 & strong core	\\		
\textit{m10zCR+AGN} & m10z	& Yes	& Yes & 1.7 &0.66 & 0.31& weak core \\
\bottomrule
\end{tabular}
\end{center}
\label{tab:dwarfruns}
\end{table*}

The modelling of BHs and CRs remains highly uncertain. Several studies have investigated the impact of different modelling assumptions for CRs \citep[e.g.][]{hopkins_but_2020,hopkins_simple_2023,butsky_constraining_2023} and BHs \citep{angles-alcazar_black_2017,angles-alcazar_cosmological_2021,su_which_2021,su_unravelling_2024,byrne_stellar_2023,byrne_effects_2023,wellons_exploring_2023} within the \textsc{fire} model. For our study, we base our BH model on the extensive study by \citet{wellons_exploring_2023} who identified a modelling space that allows for efficient BH feedback whilst still matching observational constraints, in particular with regards to star formation distributions. We summarise these BH modelling choices in Section~\ref{subsubsec:methods_bhs} and point the interested reader to \citet{wellons_exploring_2023} for more details and an in-depth parameter study of alternatives to our choices. For the CR modelling, we employ the numerically-efficient subgrid CR model introduced in \citet{hopkins_simple_2023} allowing us to explore several different IC set-ups to $z=0$ whilst minimising the computational overhead costs that would arise from explicit CR transport. We summarise the key characteristics of the CR subgrid model in Section~\ref{subsubsec:methods_crs} and refer the interested reader to \citet{hopkins_simple_2023} for a more detailed overview of this approach.

\subsubsection{BH modelling} \label{subsubsec:methods_bhs}
For our BH modelling, we largely adopt the default \textsc{fire-3} BH implementation \citep[see][]{hopkins_fire-3_2023}, except for the AGN feedback injection where we use the continuous wind mode (rather than particle spawning) and for the CR injection where we employ the subgrid model from \citet{hopkins_simple_2023}. Below we summarise the salient details as well as our modifications to the \textsc{fire-3} BH model and refer the interested reader to \citet{hopkins_fire-3_2023}  and \citet{wellons_exploring_2023} for a more in-depth description and parameter studies.

BHs may be seeded into the simulation from any star-forming gas particles that satisfy physically-motivated properties for the formation of low-mass seeds. In particular, the gas particles must have extremely high densities ($\Sigma_\mathrm{gas}\gtrsim\,5000\,\Msun\,\mathrm{pc}^{-2}$) and low metallicities ($Z_\mathrm{gas}\lesssim0.001\,\Zsun$) in order to spawn BH particles of mass $\Mbh=100\,\Msun$. With a DM particle mass of $M_\mathrm{DM} \sim 10^{3} \ \Msun$, we therefore do not fully resolve the dynamical friction forces that would drive the orbits of our seed BHs towards the centre of the galaxy. Hence we follow \citet{wellons_exploring_2023} and employ a subgrid prescription where the BHs are `drifted' towards the local binding energy extremum of the stars, dark matter and BH particles so that they are not artificially ejected \citep[also see][]{hopkins_fire-3_2023}. Similarly, since we cannot resolve the dynamics of binary BHs, two BHs are merged if their interaction kernels and force softenings overlap and they are gravitationally bound to one another.

Once a BH particle is present, it grows via accretion by draining gas from the nearest $N\sim256$ gas cell neighbours.  \citet{wellons_exploring_2023} parameterized the accretion rate as $\Mdotacc \equiv \eta_\mathrm{acc}\Mgas\Omega$, where $\eta_\mathrm{acc}$ is the accretion efficiency, $\Mgas$ is the gas mass within the BH kernel, and $\Omega = \sqrt{GM_\mathrm{tot}/R}$ is the orbital frequency determined at the kernel size $R$, based on the total mass enclosed within $R$, $M_\mathrm{tot}$. \textsc{fire-3} modulates the BH accretion rate via a gas reservoir that represents an accretion disc from which the BH particle grows at a mass growth rate $\Mdotbh = M_\mathrm{\alpha disc} / t_\mathrm{dep}$, where $M_\mathrm{\alpha disc}$ is the mass of the accretion disc and $t_\mathrm{dep} = 42~\mathrm{Myr} \ (1 + \Mbh / M_\mathrm{\alpha disc})^{0.4}$ is the depletion time.

While \citet{wellons_exploring_2023} tested various distinct accretion models (i.e. variations in $\eta_\mathrm{acc}$), their best-fitting models favoured accretion that was powered by gravitational torques. Physically, these torques result from asymmetries in a galaxy's gravitational potential, or from interactions between (a) the dark matter and gas, (b) the stellar component and gas, or (c) the self-interaction of the gas with itself via shocks and dissipation \citep{hopkins_how_2010,hopkins_analytic_2011}. The accretion efficiency in the case of gravitational torque-dominated accretion comes from the work in \citet{hopkins_analytic_2011},

\begin{equation}
    \label{eq:torque_accretion_efficiency}
    \eta_\mathrm{acc} = C \frac{([\Mbh + M_\mathrm{d}] / M_\mathrm{d})^{1/6}}{1 + 3M_\mathrm{d,9}^{1/3}(\Mgas/M_\mathrm{d})},
\end{equation}

\noindent where $C$ is a resolution-dependent constant, $\Mbh$ is the BH mass, $M_\mathrm{d}$ is the mass of angular-momentum support material, and $M_\mathrm{d,9} = M_\mathrm{d} / 10^{9}\,\Msun$.  The model has been studied extensively in the context of cosmological zoom-in simulations in \citet{angles-alcazar_black_2013,angles-alcazar_black_2017,angles-alcazar_gravitational_2017,angles-alcazar_cosmological_2021} and \citet{hopkins_stellar_2016}.  It is important to note that we choose a value of $C = 0.1$ in order to have reasonable BH masses within our three simulated dwarf galaxies.  However, the parameter $C$ cannot take into account all sub-grid processes that may occur below the resolution scale.

One important piece of (partially) unresolved physics is the impact of stellar feedback on the accretion flow in the vicinity of the BH.  \citet{wellons_exploring_2023} introduce a scaling factor $f_\mathrm{acc}$ (such that $\eta_\mathrm{acc} \to f_\mathrm{acc}\eta_\mathrm{acc}$) that depends on the local gravitational acceleration, $a_\mathrm{g,eff} = GM(<R)/R$, as 
\begin{equation}
    \label{eq:gravitational_acceleration_factor}
    f_\mathrm{acc} = \frac{a_\mathrm{g,eff}}{a_\mathrm{g,eff} + a_\mathrm{g,crit}},
\end{equation}
\noindent where $a_\mathrm{g,crit} \approx 10^{-7}\,\mathrm{cm}\,\mathrm{s}^{-2}$.  Below $a_\mathrm{g,crit}$ (an effective surface density of $\Sigma_\mathrm{tot} \sim 3000\,\Msun\,\mathrm{pc}^2$), a significant fraction of gas is expelled due to stellar feedback \citep{grudic_maximum_2019,hopkins_why_2022}. However, we emphasise that in addition to this stellar feedback effect on unresolved scales, our simulations explicitly capture the impact of stellar feedback on larger resolved scales (comparable to the size of the BH kernel).

The accretion process itself drives powerful outflows, jets and radiation that influence the properties of the galaxies hosting the BHs, collectively coined BH feedback. The \textsc{fire-3} BH feedback model includes three channels: radiative feedback, mechanical feedback, and CR feedback.  Each mode injects energy at a rate proportional to the BH accretion rate $\Mdotbh$.

As gas falls into a BH, it loses gravitational energy that is converted into radiation, with an efficiency that is dependent on the location of the innermost stable circular orbit of the BH. While the location of that orbit depends on the spin of the BH, the typical efficiency used in the literature is $\epsilon_\mathrm{r} = 0.1$, implying that 10\% of the rest mass-energy accretion rate is converted into radiation. Given that we do not track the BH spin evolution in our simulations, we also adopt this fiducial radiative efficiency value in our work. In \textsc{fire-3}, that radiation is treated in the same manner as the stellar feedback --- with multiband radiation transport and metallicity-dependent opacities using the LEBRON method \citep[see][]{hopkins_radiative_2020}, except using a quasar template spectrum \citep{shen_bolometric_2020}.  The radiation momentum flux depends on the luminosity absorbed ($L_\mathrm{abs}$) in the gas as $\dot{p} = L_\mathrm{abs} / c$.  

There are other outflows generated from the accretion process besides radiation-driven winds.  For example, there may be jets or hydromagnetic winds from the accretion disc itself. The physics that drives these processes occurs on scales that are much below the resolution of our simulations and, therefore, need to be parameterized into a sub-grid model. The \textsc{fire-3} BH model assumes that these winds have mass outflow rates proportional to the accretion rate. We assume a mass loading of one, such that $\dot{M}_\mathrm{out} = \Mdotbh$. The kinetic energy rate in the wind itself depends on the wind velocity at launch, $\vwind$, such that $\dot{E}_\mathrm{mech} = \Mdotbh\vwind^2 / 2$.  \citet{wellons_exploring_2023} recast the mechanical luminosity in terms of the new mechanical efficiency, $\eta_\mathrm{mech} \equiv \vwind^2 / (2c^2)$, such that $\dot{E}_\mathrm{mech} = \eta_\mathrm{mech}\Mdotbh c^2$.  In this work, we choose $\vwind = 10,000\,\kms$ (see e.g. \citealt{fiore_agn_2017}, compiled in fig. 3 of \citealt{rennehan_obsidian_2024}) and, hence, have $\eta_\mathrm{mech} = 5.6\times10^{-4}$. 

The \textsc{fire-3} model also includes a CR model and, therefore, it also includes a BH feedback channel that accounts for relativistic massive particles accelerated by the AGN jets. We describe the CR propagation model in Section~\ref{subsubsec:methods_crs}, but note here that the CRs are injected at a rate $\dot{E}_\mathrm{CR} = \eta_\mathrm{CR}\Mdotbh c^2$, where $\eta_\mathrm{CR}$ is the efficiency of CR feedback. In this work, we use an efficiency of $\eta_\mathrm{CR} = 0.01$ which was identified as yielding realistic galaxy properties in the \citet{wellons_exploring_2023} parameter study.

For the mechanical and radiative feedback, we use the solid angle weighting to inject the AGN feedback \citep[akin to SN feedback injection, see][for details]{hopkins_fire-2_2018}. Following \citet{su_which_2021}, we inject the majority of the AGN energy within a relatively narrow opening angle by weighting the AGN wind energy and radiation pressure received by each gas cell as
\begin{equation}
w(\theta) = \frac{\epsilon_\mathrm{jet} (\epsilon_\mathrm{jet} + \cos^2\!\theta)}{(\epsilon_\mathrm{jet}+1)(\epsilon_\mathrm{jet}+(1-\cos^2\!\theta))}.
\end{equation}
We set $\epsilon_\mathrm{jet}=0.35$ which leads to an effective base opening angle of $\sim 7 \degree$. The angle weighting is applied with respect to the total gas angular momentum accreted by the BH, $\bm{J}_\mathrm{gas}$, which is measured in the centre-of-mass frame of the BH -- accretion disc sink particle based on the kernel-weighted accreted mass and relative position and velocity of the respective gas cell neighbour. This allows for AGN energy injection without significantly disrupting the small-scale gas supply in the central region which favours repeated bursts and/or continuous AGN activity allowing us to probe the `physically interesting' regime for AGN influencing cusp to core transformations. As discussed in \citet{su_which_2021}, the large-scale collimation properties of the jet are not hugely sensitive to the small-scale opening angle since the feedback energy chooses the path of least resistance effectively recollimating even for large opening angles, also see \citet{koudmani_fast_2019}. Long-range radiation transport and CRs are treated via the LEBRON approximation and injected isotropically \citep[see][]{hopkins_radiative_2020,hopkins_simple_2023}.

\begin{figure*}
    \centering
    \includegraphics[width=0.85\textwidth]{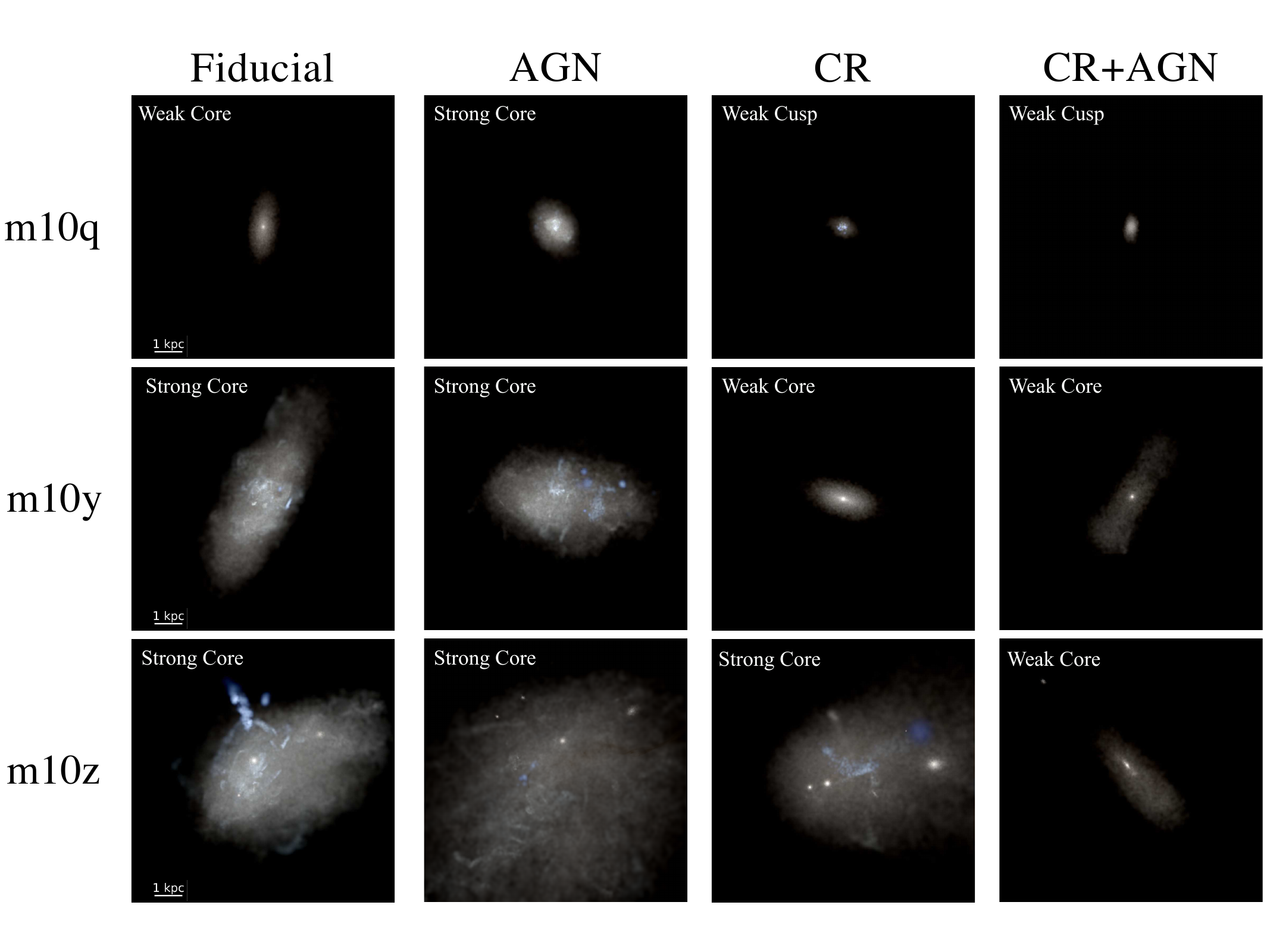}
    \caption{Mock u/g/r Hubble images showing the stellar distributions of our simulated dwarf galaxies. The side length of each projection is 10~kpc. The rows depict our three different IC set-ups, whilst the columns represent the different galaxy formation physics variations explored. The stellar sizes and colours are clearly impacted by the addition of CRs and/or AGN feedback hinting at a potential influence by these processes on dark matter distributions.}
    \label{fig:proj_suite}
\end{figure*}

\subsubsection{CR modelling} \label{subsubsec:methods_crs}

The \textsc{fire-3} model includes a comprehensive CR model that explicitly follows the multi-species/multi-spectral CR dynamics \citep{hopkins_cosmic_2021,hopkins_effects_2021,hopkins_testing_2021,hopkins_fire-3_2023} with the entire distribution function \citep{hopkins_standard_2022} integrated with the magnetohydrodynamics (MHD) solver.  However, there are also optional, simplified models that make a series of approximations valid in our regime of interest --- dwarf galaxies.  In particular, we use the sub-grid CR model described in \citep{hopkins_simple_2023} and briefly describe our motivation for this model below. 

The primary assumption is that the CR energy spectrum is spatially constant, and that CRs above $\gtrsim 1\,\mathrm{GeV}$ dominate the CR pressure \citep[e.g.][]{chan_cosmic_2019}.  This ``single-bin'' model only evolves the integrated CR energy density, drastically reducing the computational time while providing a reasonable approximation of the CR pressure, which is the dominant variable that impacts galaxy evolution.  The study in \citet{chan_cosmic_2019} showed that higher values of the isotropic diffusion coefficient $\kappa$ are necessary to explain $\gamma$-ray luminosities in dwarf and $L^*$-galaxies. Importantly, the values of the diffusion coefficient that best match the observations leads to most CRs escaping from galaxies with low gas density (i.e. dwarfs), while in starbursting galaxies with dense gas, the CR energy is lost almost entirely to collisions.  Once the CRs escape the dense gas, the adiabatic gain/loss terms from the $P\,\mathrm{d}V$ work on the low density gas also becomes negligible.

Given that several of the CR physical processes are seemingly subdominant in the dwarf regime, \citet{hopkins_simple_2023} made a series of further simplifying assumptions that are applicable in a `streaming+diffusion' limit.  In particular, the reduced sub-grid model assumes that the CR energy equation is in steady state ($\partial_t e_\mathrm{cr}\to 0$) and that the magnetic fields below the resolution scale are isotropically `tangled'. The latter assumption allows the anisotropic diffusion tensor $\boldsymbol{\kappa}_\mathrm{||}$ to be replaced with an isotropic diffusion coefficient $\kappa$, averaged over the resolution scale (and spatio-temporally constant).  The dominant loss term for CR energy is assumed to be a combination of hadronic/pionic, Coulomb, and ionization losses --- therefore, neglecting losses from diffusive reacceleration \citep{hopkins_standard_2022}.  Streaming contributions from advective and Alfv\'en velocities are also ignored, leading to a two parameter model depending on the isotropic diffusion rate $\kappa$ and a streaming velocity $v_\mathrm{\kappa}$.  An approximate solution for the spatial distribution of CR energy density for such a model is presented in \citet{hopkins_simple_2023}, and depends on an integral of an exponentially-declining function of the loss function.

The resultant CR model is solved using a LEBRON-type approximation \citep[see][for details]{hopkins_fire-2_2018}. The `optical depth' of the CRs is equivalent the loss function we describe above, and the CR energy at a given distance from a source (i.e. SNe or AGN) is solved using the gravity tree in \pkg{GIZMO}.  In \textsc{fire-3}, 10 per cent of the SN energy is converted to CRs and, as we mention above, 10 per cent of the BH luminosity is in the form of CRs. In this work, we use the values from \citet{hopkins_simple_2023}; $\kappa = 5\times10^{28}\,\mathrm{cm}^2\,\mathrm{s}^{-1}$ and $v_\mathrm{\kappa} = 20\,\mathrm{km}\,\mathrm{s}^{-1}$.
 
\section{Results} \label{sec:results}
\subsection{Overview of the simulation suite} \label{subsec:res_overview}

We begin our analysis of the simulation suite by inspecting visualisations of our simulated dwarfs' stellar distributions. Figure~\ref{fig:proj_suite} shows mock u/g/r Hubble images of the stars in each dwarf set-up, made by ray tracing through the gas: for each pixel, we sum up the luminosities of the star particles in each band along the line of sight. Dust associated with the gas elements attenuates the light, assuming a Milky Way-like attenuation curve in each band and a fixed dust-to-metals ratio. The side length of each projection is 10~kpc and the galaxies have been rotated to be viewed `face-on' as defined by the angular momentum vector of the star particles. 

The three rows correspond to our three different ICs whilst the different columns represent our four physics variations based on the \textsc{fire-3} physics modules: the fiducial \textsc{fire-3} set-up (without CRs or AGN), \textsc{fire-3} with AGN, \textsc{fire-3} with CRs and \textsc{fire-3} with both CRs and AGN. We note that there are several uncertainties associated with modelling AGN and CRs in galaxy formation simulations and here we only explore a small subset of this parameter space that was identified by \citet{wellons_exploring_2023} as producing realistic galaxy properties. Furthermore, our simulations likely represent an upper limit for the impact of CRs since we are using an effective subgrid model for the CR population \citep{hopkins_simple_2023} which only accounts for local losses and assumes a constant diffusion and effective streaming speed. See Section~\ref{subsubsec:methods_crs} and \citet{hopkins_simple_2023} for more details.

\begin{figure*}
    \centering
    \includegraphics[width=\textwidth]{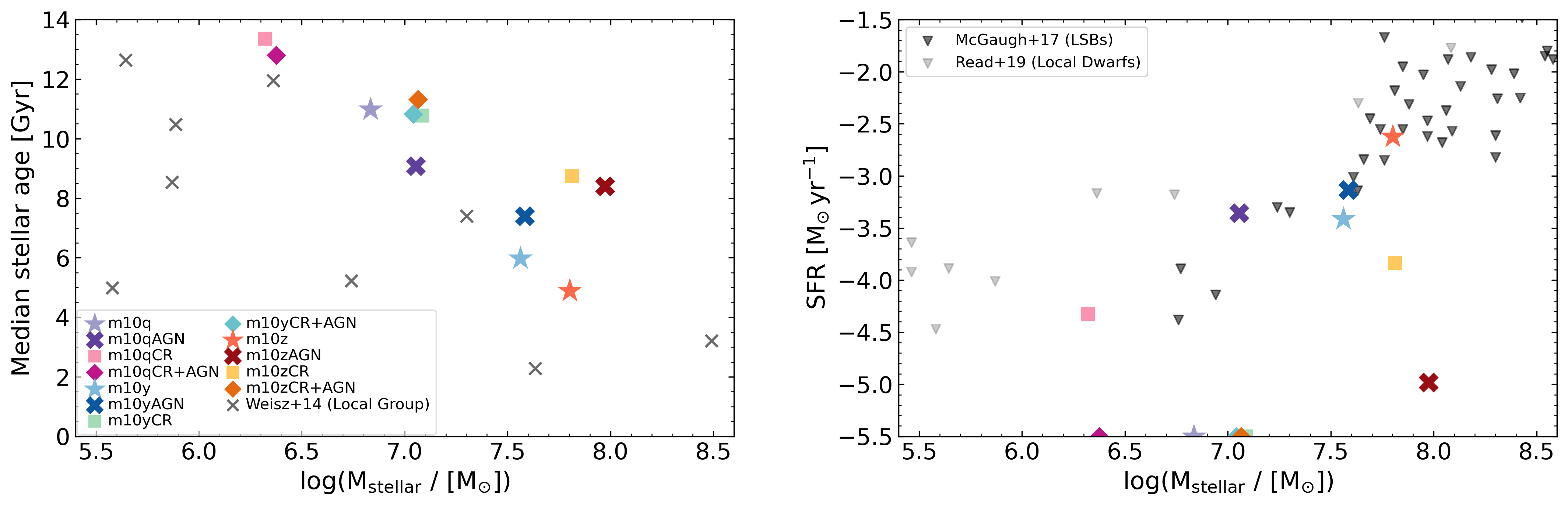}
    \caption{Star formation properties of our simulations. The symbol style in this and later plots indicates the model variations. \textit{Left panel:} Median stellar age as a function of stellar mass with galaxy stellar ages from Local Group dwarfs shown as grey crosses for comparison \citep{weisz_star_2014}. \textit{Right panel:} SFRs as a function of stellar mass. Observed SFRs for local star-forming dwarfs \citep{read_abundance_2019} and low-surface-brightness galaxies \citep{mcgaugh_star-forming_2017} are also shown for comparison as grey triangles. The inclusion of CRs leads to markedly older stellar populations and AGN-driven CR feedback results in quenched dwarfs at $z=0$ for all set-ups explored.}
    \label{fig:sfr_prop}
\end{figure*}

Keeping these caveats in mind, we note that both CRs and AGN have a significant impact on the stellar distributions. Generally, the addition of CRs leads to more compact stellar distributions, whereas the simulations without CRs also have an extended diffuse component. This is partly due to CRs suppressing star formation as the additional pressure support prevents gas accretion from the CGM onto the galaxy \citep[also see][]{hopkins_but_2020,hopkins_simple_2023,farcy_radiation-magnetohydrodynamics_2022,defelippis_effect_2024,thomas_why_2024}. For the runs with AGN feedback, the picture becomes even more complex. The runs without CRs but with AGN appear qualitatively similar to their non-AGN counterparts if not slightly more diffuse. The runs with CR, however, display systematic differences with added AGN feedback. This is consistent with the analysis from \citet{wellons_exploring_2023} where they find that CR-driven feedback is among the most efficient AGN feedback channels. In all cases, the runs with AGN are more compact and redder, pointing towards older stellar populations. Overall, these visualisations demonstrates that both AGN and CRs have a significant non-linear impact on dwarf galaxy evolution. 

\subsection{Stellar assembly} \label{subsec:res_stellar}

Before assessing the impact of our baryonic modelling choices on dark matter distributions, it is vital to confirm that our feedback implementations produce realistic dwarf galaxy properties. To this end, we assess the stellar properties of our simulated dwarfs in Figure~\ref{fig:sfr_prop}. Note that in this and later plots the symbol style indicates the model variations explored in our simulation suite, as listed in the legend. We show the median stellar age as a function of stellar mass in the left panel. The galaxy stellar ages of Local Group dwarf galaxies from \citet{weisz_star_2014} as compiled by \citet{mercado_relationship_2021} are also plotted for comparison. In the right panel, we show the SFRs of our dwarfs at $z=0$ averaged over the past $500$~Myr as well as the observed SFRs of local dwarfs \citep{read_abundance_2019} and low-surface-brightness galaxies \citep{mcgaugh_star-forming_2017}. The \qC \ and \qCA \ simulations have markedly older stellar populations, corresponding to ancient ultra-faint dwarf galaxies \citep[e.g.][]{brown_quenching_2014}, than their counterparts without CRs. The same trend holds for the other two dwarf haloes, albeit less pronounced. The SFRs of the star-forming simulated dwarf galaxies are broadly consistent with the star-forming dwarfs in the Local Group. Several of our dwarf simulations have been quenched for the past 500 Myr (indicated at the bottom of the right panel) consistent with the growing evidence for self-quenched dwarfs in low-density environments \citep{kaviraj_quenching_2025}. In particular, all simulations with AGN-driven CR feedback result in quenched dwarf galaxies at $z=0$. Finally, we note that the early formation of \qC \ and \qCA \ also means that these set-ups are very sensitive to the assumptions on primordial chemical enrichment and, since we do not account for enrichment from Pop III stars, this leads to very low metallicities for these two simulated dwarfs ($\mathrm{[Fe/H]} \lesssim -3.0$, not shown), also see discussion in \citet{wheeler_be_2019}.

\begin{figure*}
    \centering
    \includegraphics[width=\textwidth]{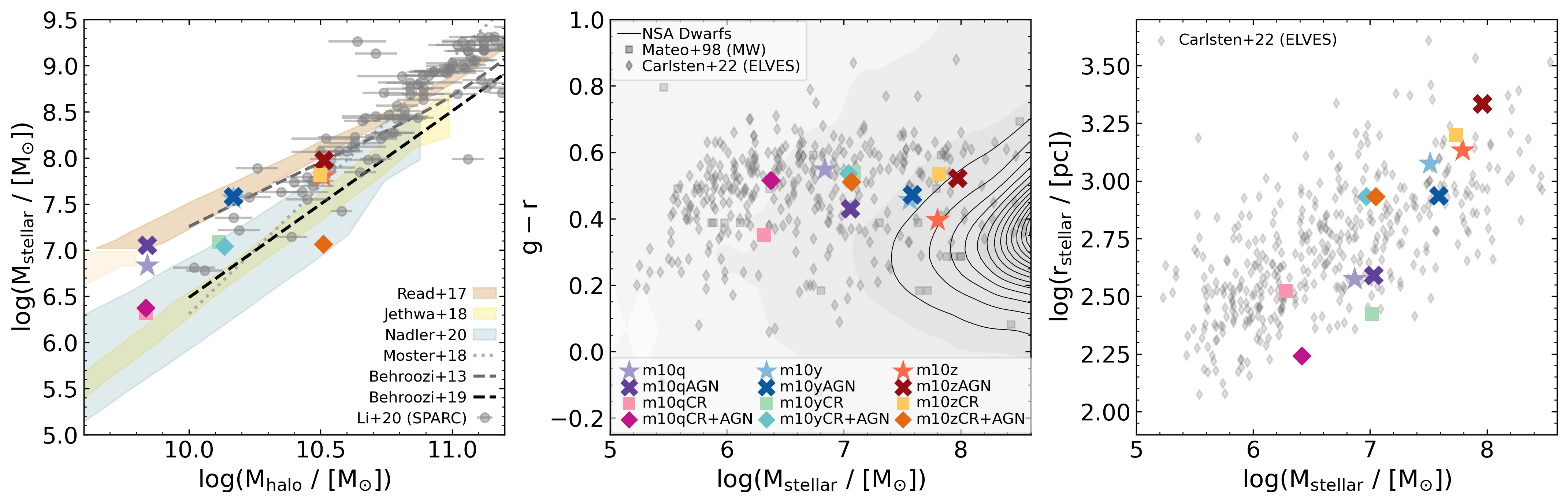}
    \caption{Integrated stellar properties of our simulations in comparison with observational constraints. \textit{Left panel:} Stellar mass -- halo mass (SMHM) relation. SMHM relations based on empirical models as well as individual observed galaxies are plotted for comparison as indicated by the legend. \textit{Middle panel:} (Intrinsic) simulated galaxy colours against stellar mass. The colours from the SDSS galaxies are indicated as grey contours and the integrated photometry of local dwarfs is plotted as grey squares \citep{mateo_dwarf_1998} and grey diamonds \citep{carlsten_exploration_2022}, for comparison. \textit{Right panel:} Stellar half-mass radii against stellar mass with the data from local dwarfs \citep{carlsten_exploration_2022} shown as grey diamonds for comparison. \textit{Overall}, though the addition of AGN and/or CRs introduces significant scatter in the simulated distributions, our simulated stellar properties are in good agreement with the observations for all simulation set-ups explored.}
    \label{fig:stellar_prop}
\end{figure*}

Motivated, by this initial exploration of stellar assembly trends, we present further comparisons with observational constraints in Fig.~\ref{fig:stellar_prop}, which shows the stellar mass -- halo mass (SMHM) relation, galaxy colours and stellar half mass radii. 

\subsubsection{SMHM relation}

Firstly, we examine if the simulated dwarf galaxies align with observational expectations for the SMHM relation. The left panel of Fig.~\ref{fig:stellar_prop} shows our 12 dwarf zoom-in simulations in $M_\mathrm{halo}$ -- $M_\mathrm{stellar}$ space at redshift $z=0$.

For comparison, we plot several empirical SMHM relations from the literature, including \citet{behroozi_average_2013, behroozi_universemachine_2019} and \citet{moster_emerge_2018}. The low-mass end of the SMHM relation remains largely unconstrained by observations, evident from the significant differences between these models. The \citet{behroozi_average_2013, behroozi_universemachine_2019} models extend down to $\log(M_\mathrm{halo} / \Msun) = 10.0$, and the \citet{moster_emerge_2018} relation was shown to also be valid down to the dwarf regime, at least to $\log(M_\mathrm{halo} / \Msun) = 10.0$, by \citet{oleary_predictions_2023}. Extrapolating SMHM relations below the minimum mass considered has pitfalls, especially regarding physical processes like reionization suppression, which are effective at low masses but not included in most of the models. The scatter is especially difficult to model in this regime since more bursty star formation histories at lower stellar masses are expected to increase the scatter whilst high-redshift quenching due to reionization suppression may act to substantially reduce the scatter in the dwarf regime \citep{oleary_predictions_2023}. Indeed, there is also evidence from various studies that at very low halo masses ($\lesssim 10^{9} \ \Msun$) the SHMH relation may converge towards a constant stellar mass \citep[e.g.][]{buck_nihao_2019,munshi_quantifying_2021,arora_nihao-lg_2022,monzon_constraining_2024}.

Our simulations with the \textbf{m10q} ICs, the least massive halo, are firmly in the extrapolated regime, whilst \textbf{m10y} and \textbf{m10z} are at the lower end of validity for the \citet{behroozi_average_2013,behroozi_universemachine_2019} and \citet{moster_emerge_2018} relations. Hence for our simulations comparing to SMHM relations constructed for low-mass galaxy samples, rather than the fiducial SMHM relations for the general galaxy population, is more instructive. We include SMHM relations derived for dwarf galaxies by \citet{read_stellar_2017}, \citet{jethwa_upper_2018}, and \citet{nadler_milky_2020} as shaded regions. The differences in these relations arise due to different assumptions for surface-brightness incompleteness correction, see discussion in \citet{behroozi_universemachine_2019}. We also show various observed data points from the SPARC data base \citep{li_comprehensive_2020} to indicate the scatter in the observations\footnote{Note that the halo masses here are derived assuming an NFW profile in the $\Lambda$CDM cosmology.}. Generally, there is always a 0.3 dex systematic uncertainty on stellar mass estimation \citep{courteau_global_2015} from observations which arises from different modelling choices with regards to e.g. star formation histories and dust models. Here the stellar masses are based on an assumed constant stellar mass-to-light ratio of $0.5 \ \Msun/\mathrm{L_{\odot}}$ at $3.6 \ \mathrm{\micro m}$ \citep[see][]{lelli_sparc_2016}.

We note that the empirical relations as well as the SPARC data base use different definitions for the halo mass, either employing the virial mass as defined by the mass enclosed in a sphere whose mean density is 200 times the critical density of the Universe at $z=0$ or the halo mass based on the collapse of a spherical top-hat perturbation following \citet{bryan_statistical_1998}. We have checked for our simulated dwarfs that the difference between these two definitions is at most 0.1 dex, and plot the simulated halo masses as the average value between these two definitions. The simulated stellar mass is calculated within twice the stellar half mass radius. Given the significant uncertainty and scatter in SMHM relations, we do not adjust the stellar masses for different radial cut-offs or initial mass functions.

From this comparison with the empirical relations and SPARC observations, none of our models can be ruled out based on their integrated stellar masses. All our simulations fall within the range of the empirical SMHM relations and are well within the scatter of the observed data. The simulation set-ups with CRs (indicated by squares and diamonds) are generally less efficient at forming stars than the simulations without CRs (indicated by stars and crosses). Most simulations fall towards the upper end of the predicted SMHM relations, with the \zCA \ simulation the only set-up resulting in an `undermassive' galaxy. Though we note that even this set-up is still within the scatter of the most-up-to-date dwarf galaxy SMHM relation by \citet{nadler_milky_2020}.

\subsubsection{Galaxy colours}

Next, we compare the galaxy colours of our simulated dwarfs with observational data. The middle panel of Fig.~\ref{fig:stellar_prop} presents the $g - r$ colour versus stellar mass for our simulated dwarfs. For comparison, we display the colours of SDSS galaxies from the NASA Sloan Atlas (NSA) as a grey contours, the colours of Milky-Way dwarf satellites from \citet{mateo_dwarf_1998} as grey squares and the ELVES dwarf satellites, a nearly volume-limited sample of Milky Way-like hosts in the Local Volume, from \citet{carlsten_exploration_2022} as grey diamonds.

We use the \citet{bruzual_stellar_2003} stellar population synthesis model to calculate galaxy colours as a function of stellar age and metallicity, assuming a \citet{kroupa_variation_2001} IMF, including all star particles within twice the stellar half-mass radius. For the MW dwarf satellites, we convert the $B - V$ colours using the transformation equations from \citet{jester_sloan_2005}, noting that while these equations are for stars, they should approximate galaxies well unless strong emission lines are present. One important caveat is that our simulated galaxy colours do not account for dust attenuation. However, we checked that, as expected for dwarf galaxies, our simulations have relatively low metallicities and therefore dust attenuation is expected to be minimal.

All simulated dwarfs are overall in good agreement with both the local dwarf galaxy constraints and the colour distribution of SDSS galaxies. Although we note that SDSS is very incomplete in the mass regime we consider here and the ELVES dwarfs were selected to be satellites while our dwarfs are field objects. The SDSS galaxies cluster around $g-r\sim0.35$, whilst most of our dwarfs are offset towards somewhat redder colours, clustering around $g-r\sim0.45$. Accounting for dust attenuation would likely make this offset more severe. However, this discrepancy is also inherently linked with completeness bounds in SDSS in this mass range, with blue galaxies likely overrepresented \citep{kaviraj_quenching_2025}.

Interestingly, the \qC \ set-up has a similar stellar mass to its AGN counterpart \qCA, but is shifted towards bluer colours due to a late-time burst in star formation which is also visible in the stellar projections in Figure~\ref{fig:proj_suite}. Conversely, the \qF \ set-up has a lower stellar mass and is redder than its AGN counterpart, indicating a complex interplay between CRs and AGN activity. For \textbf{m10y}, neither CRs nor the AGN have a strong impact on the integrated galaxy colours, though there are small, yet systematic, offsets with the CR runs being redder than the no CR runs, and the AGN runs being redder than their counterparts without AGN. These trends can also be seen in the median stellar ages in Figure~\ref{fig:sfr_prop}. Finally, for \textbf{m10z}, the set-up with neither CRs nor AGN yields the bluest galaxy colour, \zF, whilst the three other set-ups all have relatively similar colours, despite spanning an order of magnitude in stellar mass, indicating that all three of these formed the majority of their stars earlier in cosmic history as seen in the left-hand panel of Figure~\ref{fig:sfr_prop}. 

\subsubsection{Stellar sizes}

We now turn towards examining the stellar sizes which provide crucial clues to both the star formation and dynamical histories of our dwarfs. We plot the stellar half mass radii of our simulated dwarfs at $z=0$ as colour-coded symbols. For comparison, we also plot the observed effective radii of local dwarf galaxies as grey diamonds based on the ELVES catalogue from \citet{carlsten_exploration_2022}, which also includes the Milky Way dwarf measurements from \citet{mcconnachie_observed_2012} as a subsample. To make this comparison more consistent, we calculate the simulated stellar half mass radii based on the averages of 2D projections along ten random sightlines with the components of the random direction vectors drawn from the normal distribution. Again there is appreciable scatter in the observations and our simulated dwarfs fall well within this scatter. 

As we found from our visual inspection of the stellar distributions in Fig.~\ref{fig:proj_suite}, the stellar distributions are generally more compact with CRs. The impact of AGN feedback on the stellar sizes is more complex, leading to both more extended distributions (e.g., compare \yC \ with \yCA) or more compact distributions (e.g., compare \qC \ with \qCA). The differences between the AGN and no-AGN set-ups are more pronounced if CRs are present, in agreement with the findings from \citet{wellons_exploring_2023} that CRs enhance the efficacy of the AGN feedback in \textsc{fire}.

Overall, our stellar sizes are in good agreement with the observations, though the \qCA  simulation is slightly more compact than local dwarfs at similar stellar masses. However, we note that the observed radii are based on stellar light whilst we are calculating the simulated radii based on the stellar mass distribution. Observational effects can add significant scatter leading to both more compact half-light radii \citep[e.g.][]{klein_size-mass_2024} or more extended observed distributions \citep[e.g.][]{parsotan_realistic_2021}, with the stellar sizes also being significantly dependent on the wavelength \citep[e.g.][]{cochrane_impact_2023}.

\subsection{BH assembly} \label{subsec:res_bhs}

Having analysed the stellar properties, we turn to analyse the evolution of the BHs in our simulations in the context of observational constraints.

\begin{figure}
    \centering
    \includegraphics[width=\columnwidth]{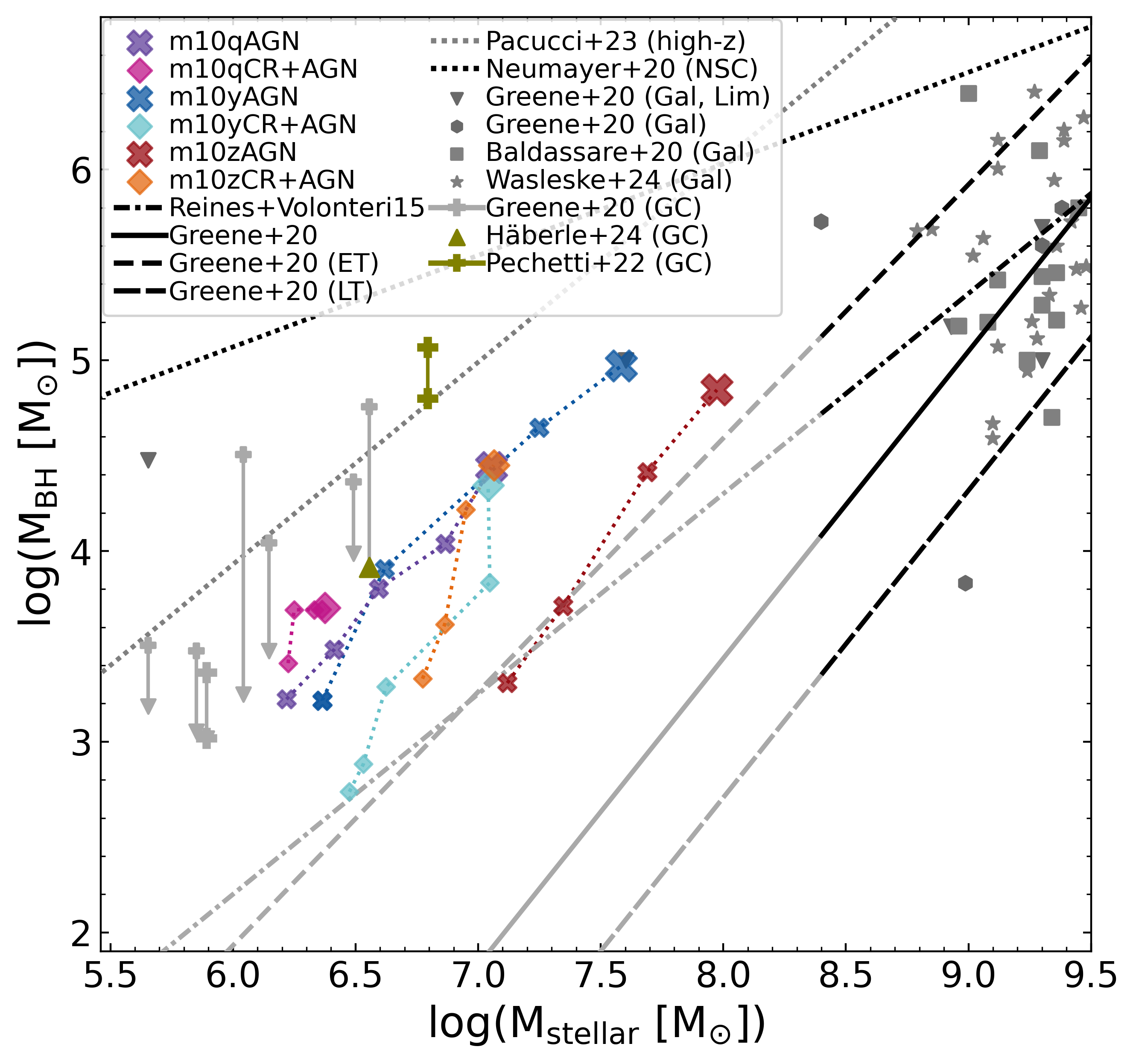}
    \caption{BH mass -- stellar mass scaling relations compared to observational constraints. We show the redshift evolution of the dwarf zoom-in simulations from $z=4$ to $z=0$ and plot data from observed dwarfs and globular clusters as indicated by the figure legends. We also show the observational BH mass -- stellar scaling relations from the literature by \citet{reines_relations_2015} and \citet{greene_intermediate-mass_2020}, greyed out in the extrapolated regime, as well as the high-redshift relation from \citet{pacucci_jwst_2023} and NSC mass -- stellar mass relation from \citet{neumayer_nuclear_2020}, which can be taken as an upper limit. Our simulated BHs are generally overmassive compared to the (heavily) extrapolated local relations though they are well within the limits set by the NSC relation and their offsets are much smaller than those uncovered by JWST at high redshifts.}
    \label{fig:scal_rel}
\end{figure}

Fig.~\ref{fig:scal_rel} shows our simulated dwarfs in stellar mass -- BH mass space. The evolutionary tracks of the dwarf simulations from $z=4$ to $z=0$ are plotted as dotted lines, colour-coded by simulation setup. Note that the BHs are generally seeded around $z \sim 8$--$10$ with the exact seeding time depending on the metallicity and density conditions. However, we here focus on later redshifts as by $z=4$, the most massive progenitor dwarf hosts a BH for all model variations so that we can track these alongside the stellar mass growth of the main progenitor. Integer redshifts are marked with the marker symbols for the respective simulation set-ups as indicated in the legend.

We note that there is substantial uncertainty at the low-mass end of the BH -- galaxy scaling relations due to the difficulty in reliably measuring the masses of intermediate-mass BHs (IMBHs) which have a much smaller gravitational sphere of influence and generally lower luminosities. Indeed there are no BH mass measurements for dwarf galaxies in the stellar mass range covered by our simulations, $10^{6} \ \Msun \lesssim M_\mathrm{stellar} \lesssim 10^{8} \ \Msun$. For more massive dwarfs there are only very limited measurements and we show these observed BH masses in massive dwarfs for reference, including virial BH mass estimates for active dwarf galaxies from \citet{baldassare_populating_2020} and \citet{wasleske_active_2024} as dark-grey squares and stars, respectively, and measurements from \citet{greene_intermediate-mass_2020} (based on dynamics, reverberation mapping or scaling from the stellar velocity dispersion in the specific case of tidal disruption events (TDEs) that seem to have reliable determinations of the host galaxies) as dark-grey hexagons and triangles for mean values and upper limits, respectively. Error bars are omitted for clarity. We note that there are no dynamical nuclear IMBH mass measurements below $M_\mathrm{stellar} \sim 10^{9} \ \Msun$ and, excluding upper limits, the lowest BH mass estimate based on TDEs stems from a host galaxy of stellar mass $M_\mathrm{stellar} = 2.5 \times 10^{8} \ \Msun$. Below this mass, we are firmly in the extrapolated regime and indicate this by greying out the local BH -- galaxy scaling relations below this mass.

To extend our comparison to lower-mass systems, we include IMBH candidates from globular clusters around the Milky Way and M31, compiled by \citet{greene_intermediate-mass_2020}, plotted in light-grey. Some globular clusters, like $\omega$ Cen, are hypothesised to be remnants of dwarf galaxies disrupted by the Milky Way's tidal field \citep[e.g.][]{freeman_globular_1993,bekki_formation_2003,meza_accretion_2005}. The presence of IMBHs in globular clusters is controversial, with significant uncertainty in BH mass values. Thus, we show a range for each cluster, indicating the lowest and highest BH mass values reported in the literature, with crosses for mean values and triangles for lower/upper limits. In addition, we plot an updated firm lower limit for the BH mass in $\omega$ Cen which was recently determined by \citet{haberle_fast-moving_2024} and a new IMBH detection in M31's most massive globular cluster which may also originate from a stripped dwarf galaxy \citep{pechetti_detection_2022}, both indicated in olive-green.

We also show several observed BH mass -- stellar mass scaling relations from the literature to give an indication of the scatter in the extrapolated relations. The \citet{greene_intermediate-mass_2020} relations are based on dynamical measurements by \citet{kormendy_coevolution_2013} as well as more recent dynamical measurements and upper limits in the low-mass regime. We plot the relation for the total galaxy sample as a solid black line, and the early-type and late-type galaxy relations as loosely and densely dashed black lines, respectively. The late-type relation has a significantly lower normalization, likely due to a weaker correlation between disc mass and BH mass \citep[e.g.][]{kormendy_inward_1995}. We also show the \citet{reines_relations_2015} scaling relation, based on local AGN, including dwarfs, as a dashed-dotted black line. This relation is significantly shallower than that of \citet{greene_intermediate-mass_2020}, leading to larger BH masses in the low-mass dwarf regime. 

For the classical dwarf mass range covered by our simulations, $M_\mathrm{stellar} \lesssim 10^{8} \ \Msun $, we would expect a flattening of the scaling relations, with the transition mass dependent on the seeding mechanism \citep[see e.g.][]{greene_intermediate-mass_2020}, which sets a lower limit for BH masses. The extrapolated BH -- galaxy scaling relations may hence be considered lower limits for BH masses in the dwarf regime. We also plot the high-redshift scaling relation derived by \citet{pacucci_jwst_2023} from overmassive BHs uncovered by JWST to indicate the high BH mass to stellar mass ratios that can be reached with optimal growth conditions in the early Universe (modulo observational uncertainties). To estimate upper limits, we show the nuclear star cluster (NSC) -- galaxy scaling relations. NSCs and massive BHs often coexist in galaxies, with NSCs being the dominant central mass component in dwarfs\footnote{The scatter for the $M_\mathrm{BH}/M_\mathrm{NSC}$ ratio at a given galaxy mass spans over three orders of magnitude, showing a trend of increasing ratio with galaxy mass, and BHs only start to dominate at stellar masses above a few times $10^{10} \ \Msun$.}. \citet{neumayer_nuclear_2020} compile NSC mass measurements extending to low-mass dwarfs with stellar masses $\lesssim 10^{6} \ \Msun$, making the relation valid for the entire $z=0$ galaxy stellar mass range considered here.

\begin{figure*}
    \centering
    \includegraphics[width=\textwidth]{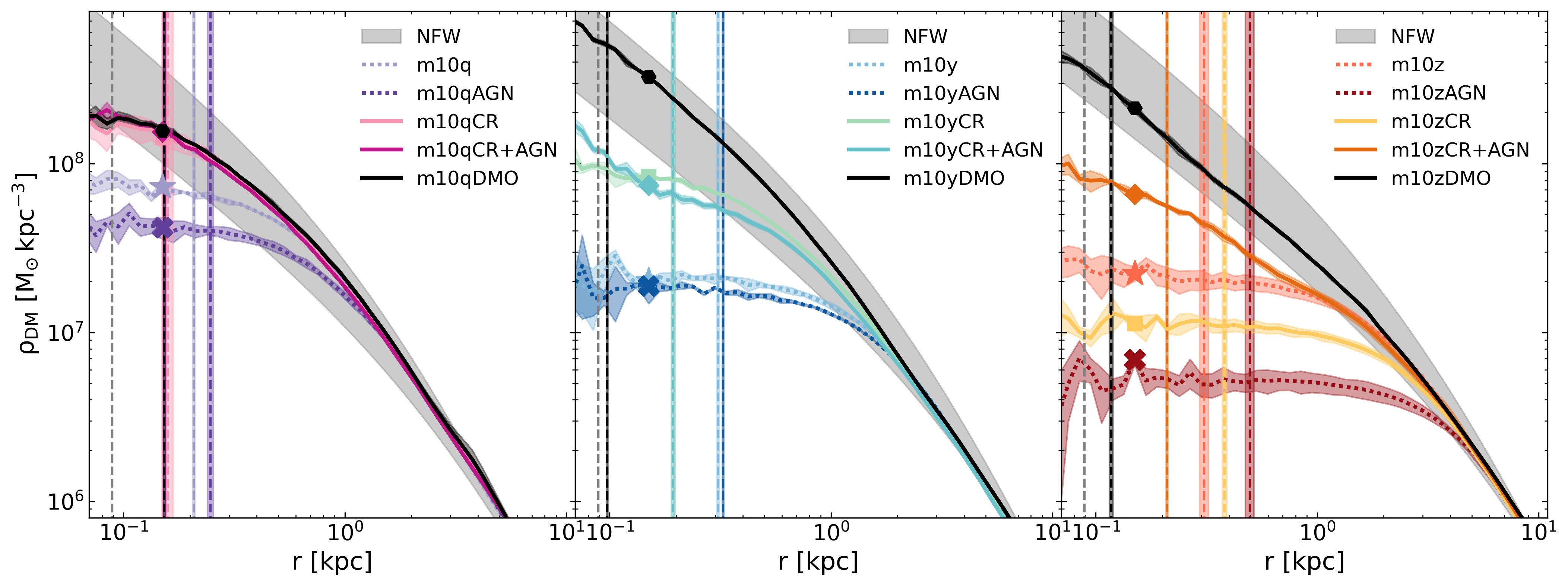}
    \caption{Spherically averaged dark matter density profiles as a function of radius. The panels display simulated profiles based on our three sets of ICs, with colour-coding indicating the galaxy formation model variations as listed in the legend. The runs without CRs are shown as dotted lines for clarity.  Density profiles from dark-matter-only (DMO) simulations are shown as solid black lines. The simulated profiles are averaged over $\Delta T=500$~Myr and the $1\sigma$ temporal scatter is shown by the colour-coded shaded regions. The expected profiles for an NFW halo are plotted as grey-shaded regions indicating the 2$\sigma$ scatter due to varying halo concentrations. The softening length for BH and dark matter particles is marked as a grey dashed line, and the Power radius  \citep[enclosing the central 2000 dark matter particles, see][]{power_inner_2003} is indicated by colour-coded dashed lines for each of the simulation set-ups. Flattening of the profiles within these radii may be partly due to numerical artefacts. CRs and/or AGN feedback cause significant diversity in dark matter profiles, ranging from cuspy to various degrees of cored profiles. Central densities at $r=150$~pc, a characteristic radius for distinguishing cusps from cores in observations \citep{read_dark_2019}, are highlighted by marker symbols corresponding to the different galaxy formation physics configurations (with the hexagon symbol representing the DMO simulations).}
    \label{fig:dm_prof}
\end{figure*}

Keeping these observational caveats in mind, we find that all of our dwarf AGN end up being overmassive compared to the extrapolated local BH -- galaxy scaling relations. Most runs are offset by about an order of magnitude with the exception of \zA \ which closely follows the early-type relation from \citet{greene_intermediate-mass_2020} and is only minimally offset from the \citet{reines_relations_2015} relation. Again, we emphasise that these offsets are not necessarily unexpected given that we are in the heavily extrapolated regime for these relations. Indeed the offsets are not as severe as the intrinsic high-redshift offsets inferred from recent JWST observations by \citet{pacucci_jwst_2023} and well within the bounds provided by the nuclear star cluster relation. 

For the redshift evolution, we plot the BH mass of the most massive progenitor for redshifts $z=1,2,3,4$. We note that in all cases the set-ups without CRs have more significant stellar and BH mass growth due to the weaker AGN feedback. The difference is especially significant at low redshifts. At late times, between $z=1$ and $z=0$, all simulations with CRs only have minimal stellar mass growth.  For \yCA \ and \zCA, there is still significant BH growth during this time indicating that whilst the CR pressure has shut down star formation, the BH is still able to accrete. For \qCA \ BH growth is already shut down from $z=3$, however, between $z=4$ and $z=3$, this run similarly experiences a rapid BH growth phase whilst star formation is already quenched.

We also investigated the role of BH mergers versus accretion in growing the BHs in our dwarf simulations and find that in all cases BH growth is predominantly driven by gas accretion. In particular, we verified that merging with other BH seeds represents a negligible growth channel with a maximum of two, five and nine seed BHs forming in the \textbf{m10q}, \textbf{m10y} and \textbf{m10z} simulations, respectively. With a seed mass of $100~\Msun$, this represents only a very small fraction of the final BH mass in all cases.

\subsection{Dark matter distributions} \label{subsec:res_dm}

Fig.~\ref{fig:dm_prof} shows the spherically averaged dark matter density profiles as a function of radius. The three panels display these profiles for our three sets of ICs, respectively. For these profiles, we centre the dark matter distributions on the minimum potential of the main halo\footnote{Note that we also investigated alternative centering methods, including the shrinking sphere method from \citet{power_inner_2003}, however, we find that in practice the differences are minimal as the common halo centre definitions are in good agreement for our dwarfs.}. In each case, we plot the mean $z=0$ profiles, averaged over $\Delta T=500$~Myr, for our four different galaxy formation model variations as colour-coded solid lines as indicated by the legend. We also performed additional dark-matter only (DMO) runs and we plot these as solid black lines. In each cases the temporal dispersion of these profiles is indicated by shaded regions which show the $1\sigma$ scatter in the densities over $\Delta T=500$~Myr. We note that for the majority of the simulation runs, the central density is not very sensitive to the temporal bin width as the evolution of the central density is relatively steady over the last Gyr. The \zC \ simulation, however, experiences significant fluctuations in the central density from $\sim 12$--$13$~Gyr due to mergers with substructures, so for our $z=0$ profiles we choose the bin width to be small enough as to not be affected by these transient features. 

For all of the simulated profiles, we highlight the central dark matter density at $r=150$~pc, a characteristic radius for distinguishing cusps from cores in observations \citep{read_dark_2019}, with colour-coded symbols. We caution when interpreting the densities at 150~pc that these are measured at a comparable scale to the gravitational softening of the dark matter and BH particles and therefore numerical effects may also impact the densities at these scales. For reference, we therefore indicate the gravitational softening length at $r\sim89$~pc as a grey dashed line. To indicate the radii outside of which the impact of numerical heating should be minimised, we also show the Power radius\footnote{We note that the required particle number for resolving the central region depends on the radius and density contrast in question but generally ranges between $N=1000$--3000; here we follow \citet{gutcke_lyra_2021} and set $N=2000$.} \citep[enclosing the nearest 2000 neighbours, see][]{power_inner_2003} for each simulation as colour-coded dashed vertical lines. We note, however, that strong temporal density fluctuations, indicative of numerical heating, are generally only seen at scales smaller than 150~pc so that we also include these scales in our analysis as a useful comparison point with observational work.

In addition, we show the expected NFW profile based on the virial mass as grey shaded regions indicating the $2\sigma$ scatter from the halo mass -- concentration relation \citep{dutton_cold_2014} to indicate the typical variations in profiles due to varying halo concentrations. The DMO runs are in good agreement with the NFW expectations, with \textbf{m10y} having slightly higher and \textbf{m10z} having slightly lower concentration than the mean relation.
 
First of all, we find that the introduction of AGN and CRs introduces a significant amount of scatter. The simulations with the default \textsc{fire-3} model, without CRs or AGN, all lead to distinct cores at $r=150$~pc. The addition of CRs and/or AGN, however, leads to a variety of outcomes from stronger cores to weakened cores to cusps at the same inner radius\footnote{We note that the central densities at $r=150$~pc are within the Power radius in most cases, however, for all cases, the cusp versus core categorisations based on the $2\sigma$ scatter around NFW expectations are unchanged between $r=150$~pc and the Power radius, demonstrating that the qualitative results are robust.}.

For simulations without CRs, the addition of an AGN enhances the core size and further suppresses the central density for all three cases. Though it should be noted that the magnitude of the additional density suppression varies and is significantly weaker for \textbf{m10y} at $r=150$~pc.

For simulations with CRs but without AGN, the profiles become cuspier than their no-CRs counterpart for \textbf{m10q} and \textbf{m10y}, whilst for \textbf{m10z} there is small suppression of the central densities with CRs. The addition of an AGN to these CR-based set-ups leads to a further increase of the central densities for \textbf{m10q}, whilst the central densities for \textbf{m10y} are minimally decreased. For \textbf{m10z} there is a much stronger impact, with the central density notable enhanced, yet still in the core regime. 

\begin{figure*}
    \centering
    \includegraphics[width=\textwidth]{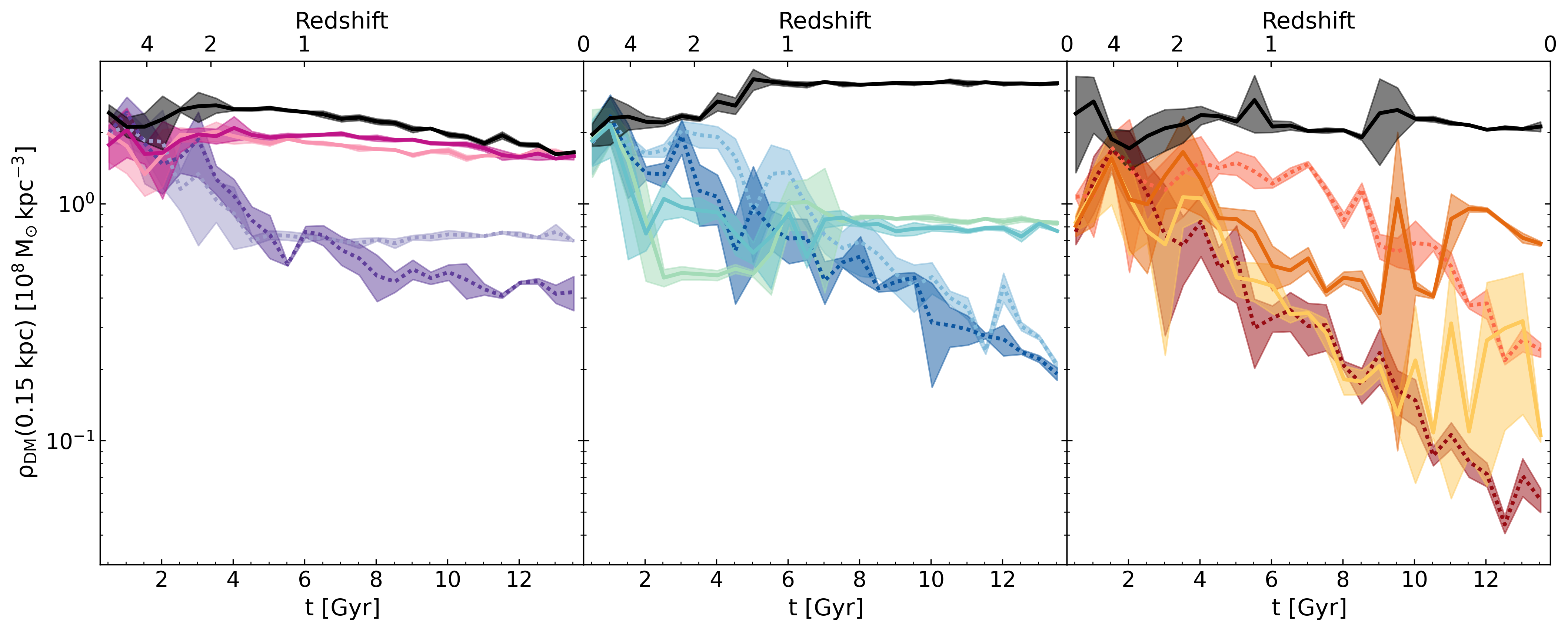} \\
    \includegraphics[width=\textwidth]{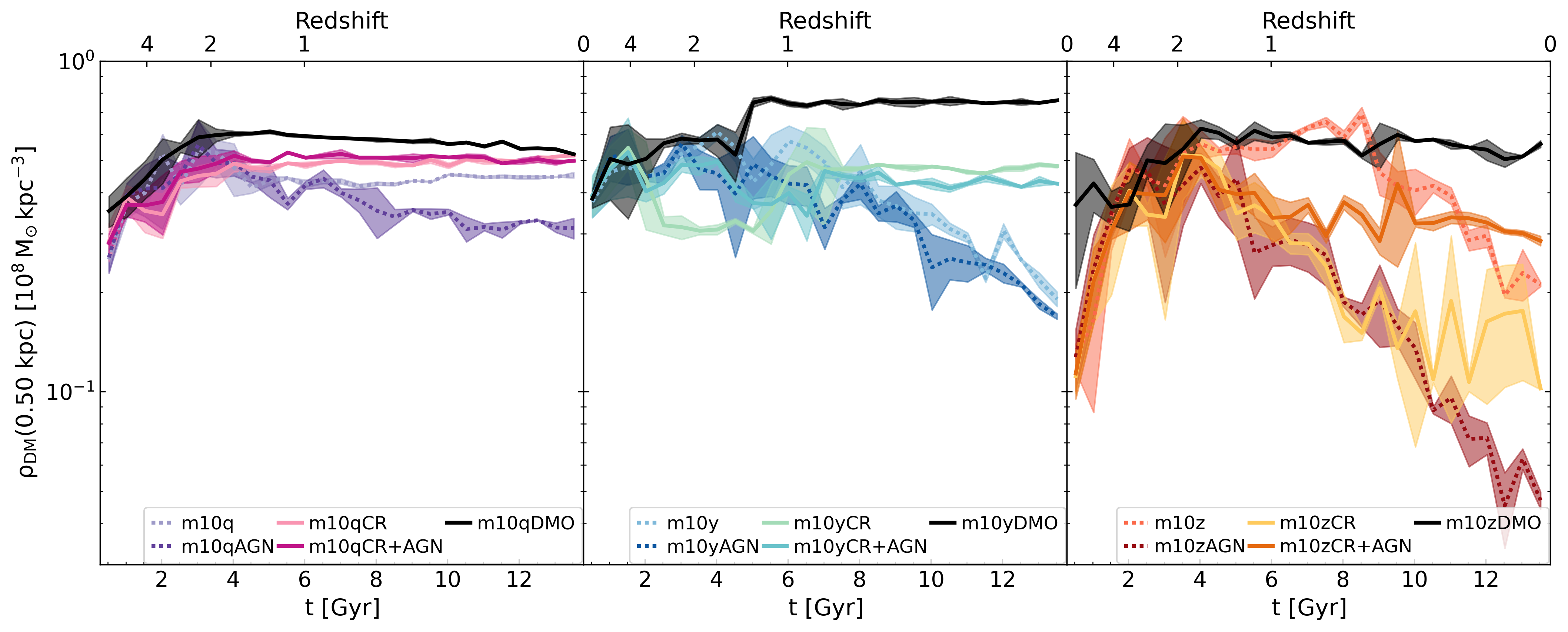}
    \caption{Central dark matter densities at $r=150$~pc (top panel) and $r=500$~pc (bottom panel) as a function of cosmic time. The dark matter densities are binned over $\Delta T = 500$~Myr and the temporal $1\sigma$ scatter in each bin is shown by the shaded regions. The colour-coding indicates the respective simulation set-ups, as listed in the legend. For clarity, we plot the simulations without CRs as dotted lines. For the \textbf{m10q} and \textbf{m10y} runs, the central densities are relatively steady at low redshifts, whilst the \textbf{m10z} runs still shows significant fluctuations due to a higher number of mergers and strong feedback events at late times.}
    \label{fig:central_dm_vs_time}
\end{figure*}

Overall, there is a variety of outcomes when introducing CRs and/or AGN to dwarf simulations which mostly do not follow clear trends due to the dual role of these processes as an additional source of feedback that may both suppress star formation (and therefore suppress core formation) or enhance core formation via the additional energy input. Hence great care has to be taken to untangle the contribution from these additional baryonic processes and their interplay with star formation. A further consideration is that small perturbations in hydrodynamical simulations (e.g. floating-point round-off or random number generators) can lead to chaotic-like behaviour with the small perturbations growing over time and manifesting as macroscopic differences in galaxy properties \citep[e.g.][]{genel_quantification_2019,keller_chaos_2019,borrow_impact_2023}. However, the variability we observe here is significantly larger than the variability expected from this `numerical butterfly effect' which typically leads to central density fluctuations of at most 10 to 15 per cent \citep[see][]{keller_chaos_2019} whereas we observe variations ranging from 70 to 90 per cent for our different sets of ICs and galaxy formation physics configurations (see Table~\ref{tab:dwarfruns}). Furthermore, as we show in the next Section, these central density trends are relatively steady as a function of cosmic time which would not be expected if the variability resulted from the amplification of small numerical perturbations. Overall, this then points to a physical origin of the diversity of dark matter profiles that we observe in our simulations.

\subsection{Driving forces of core formation} \label{subsec:res_dmforces}

Our analysis from Section~\ref{subsec:res_dm} indicates that our different feedback physics configurations significantly affect the central dark matter distributions of dwarf galaxies. In this Section, we explore the driving forces behind these trends by contrasting the cosmic evolution of the central dark matter densities with SN and AGN energy injection histories. 

Fig.~\ref{fig:central_dm_vs_time} shows the central dark matter densities at $r=150$~pc (top panel) and $r=500$~pc (bottom panel) as a function of cosmic time. In addition to our fiducial $r=150$~pc radius, an important reference point for observations, we also present the density evolution at $r=500$~pc which is at least the size of the Power radius for all set-ups explored. Again we average the dark matter densities over time bins of $\Delta T=500$~Myr and indicate the mean densities as solid and dotted lines and the standard deviation in each bin by the shaded regions. 

The \textbf{m10q} set-up is shown in the first panel. Reflecting its early formation history, the dark matter density trends we found at $z=0$ in Fig.~\ref{fig:dm_prof} have persisted for the past 6 Gyr. The central densities are relatively steady with only \qA \ showing significant evolution after redshift $z=1$ with the central density of this run showing a slow decline.

Similarly for \textbf{m10y} (middle panel), the $z=0$ trends have persisted for the past $\sim5$~Gyr. Both of the CR-based runs, \yC \ and \yCA, do not show any significant evolution at low redshifts, whilst the runs without CRs, \yF \ and \yA \ show a steady decline from $z\sim1$ onwards. Note that the density fluctuations for these runs are also much stronger, reflecting the potential perturbations characteristic of core formation \citep[e.g.][]{pontzen_how_2012,martizzi_cusp-core_2013}. Focusing on early times between redshifts $z=4$ and $z=1$, we find that most of the runs are quite similar, except for \yC \ which exhibits a strong core during that period. We also note that at $t \sim 5$~Gyr, the \textbf{m10y} system experiences a major merger. For the DMO run this significantly raises the central density, and the strong early core for the \yC \ set-up is erased by this merger.

For the \textbf{m10z} system (right panel), there is significantly more late-time evolution as well as stronger fluctuations as reflected in its more irregular morphology, see Fig.~\ref{fig:proj_suite}. Indeed, the general central density trends we observe at $z=0$ have only been in place for the past $\sim2$~Gyr and the \zC \ run experiences strong fluctuations for most of this period. We have verified that these potential fluctuations are driven by late-time mergers with substructures that pass through the centre and are independent of the slice width chosen or the method used to determine the halo centre when calculating the central density. Between redshifts $z=2$ and $z=0.5$, the set-ups \zA, \zC, and \zCA, all show a relatively similar redshift evolution of their central densities. The \zF \ set-up, however, closely follows the DMO run and, at $r=500$~pc, even surpasses it from $t=7$--$9$~Gyr, indicating strong baryonic cooling at the centre. At $z\sim 0.5$ a merger induces density fluctuations for all of our \textbf{m10z} set-ups, but most prominently for the CR-based runs \zC \ and \zCA. After this merger, the two CR-based set-ups have relatively steady central densities. On the other hand, the runs without CRs, \zF \ and \zA \ experience significant density suppressions at late times.

\begin{figure*}
    \centering
    \includegraphics[width=\textwidth]{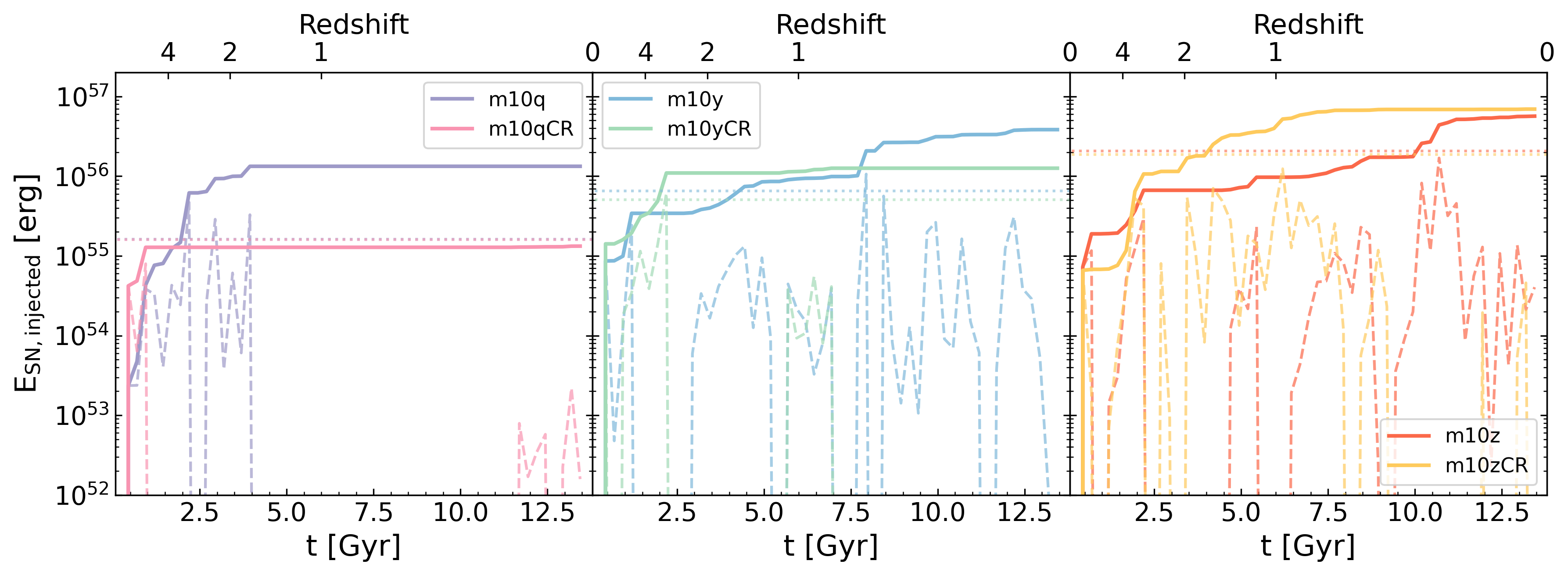}\\
    \includegraphics[width=\textwidth]{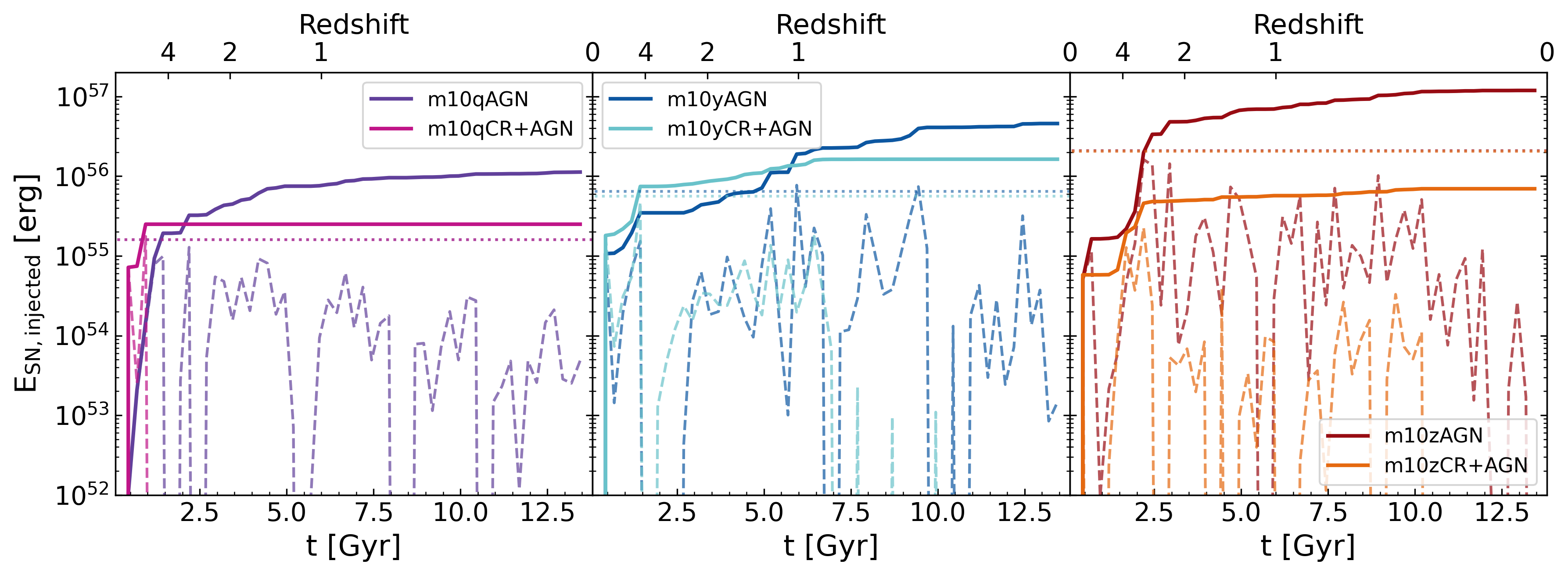}
    \caption{Instantaneous (dashed lines) and cumulative (solid lines) SN energy injected as function of cosmic time. The three columns display simulations based on our three sets of ICs, with colour-coding indicating the galaxy formation physics configurations as listed in the legends. We also indicate the $z=0$ binding energy of the halo as dotted lines. The cumulative SN energy injected compared to the binding energy is a strong predictor of core formation as well as the central dark matter density more generally. Some deviations from this pattern are introduced due to the important role of late-time star formation bursts in preserving cores.}
    \label{fig:sn_egy}
\end{figure*}

In the following two subsections, we turn to investigate the origin of these cosmic evolution patterns in the dark matter central densities, examining SN and AGN energy injection as a function of time.

\subsubsection{SN feedback} \label{subsubsec:SN_driver}

Fig.~\ref{fig:sn_egy} shows the injected SN energy as a function of cosmic time with the instantaneous energy shown as dashed lines and the cumulative energy shown as solid lines, both binned over 250~Myr for clarity. The three columns correspond to our three sets of ICs. The no-AGN and AGN runs are shown in the top and bottom row, respectively. 

We focus on the SN energy injected in the central region of the halo within a fixed 1~kpc aperture (in physical coordinates). This choice is motivated by the fiducial inner radius considered by observers for determining the shapes of dwarf galaxy rotation curves (see Section~\ref{subsec:res_diversity}) and it also corresponds to the scale of the half-mass radius for most of our systems (see Fig.~\ref{fig:stellar_prop}).

The SN energy injection rates are recalculated from the stellar ages and metallicities based on \textsc{Starburst99} tables for a \citet{kroupa_variation_2001} IMF. We focus here on the SN II rates, which constitute the majority of SN events \citep[see][]{hopkins_fire-3_2023}, for simplicity. For reference, we also indicate the $z=0$ binding energy of the respective haloes as dotted lines, which we estimate as $f_\mathrm{bar} \, M_\mathrm{vir} \, V_\mathrm{vir}^{2}$, where $f_\mathrm{bar}$ is the universal baryon fraction and $V_\mathrm{vir}$ is the virial velocity of the halo \citep[also see][]{wellons_exploring_2023}.

Firstly, we note that both CRs and AGN have a significant impact on star formation, and therefore the SN energy histories. As discussed in previous works \citep{sparre_starbursts_2017, su_discrete_2018,genel_quantification_2019,keller_chaos_2019,hopkins_what_2023}, due to their shallow potential wells, dwarf systems are very sensitive to baryonic feedback processes so that even small changes can significantly affect star formation histories.

This highly variable behaviour is quite clearly illustrated by the \textbf{m10q} set-up in the first column. With CRs, the system experiences a higher initial burst which is powerful enough to quench the dwarf system from $z \sim 6$. Hence both \qC \ and \qCA \ end up with cuspy profiles, even though the cumulative injected SN energy for \qCA \ slighly exceeds the $z=0$ binding energy, as the absence of late-time star formation means that a cusp can be re-established quite easily after the initial burst. Note that \qC \ has a slighly reduced central density at $z=0$ compared to \qCA \ due to low-level late-time star formation activity; however, these bursts are not powerful enough to induce a core. Without CRs, the initial star formation burst is less powerful, allowing for continued star formation activity. For the \qF \ set-up, there are a few more powerful bursts at intermediate redshifts, which then quench the systems at much later times from $z\sim 1.5$. For the \qA \ set-up, the AGN regulation means that the intermediate redshift star formation bursts are less intense so that star formation can continue until $z=0$. Both \qF \ and \qA \ end up with cumulative SN energy almost an order of magnitude above the $z=0$ binding energy leading to cored profiles; however, \qA \ has a more extreme core due to the continued star formation activity \citep[also see][]{muni_dark_2024}.

For \textbf{m10y}, there is a similar picture albeit not as extreme as for the \textbf{m10q} halo. With CRs, high-redshift star formation is somewhat burstier, and the more powerful bursts then also mean that most of the SN activity is restricted to $z \gtrsim 1$. This is still sufficient to (cumulatively) exceed the $z=0$ binding energy and induce a core for both of the CR-based runs; however, the cores are significantly weaker and less extended than for the runs without CRs. Both \yF \ and \yA \ have continued star formation activity until $z=0$, higher cumulative injected SN energies, and therefore much stronger cores. Adding in AGN only has a small impact for both \yA \ and \yCA \ compared to their no-AGN counterparts. Again the bursts at high redshift are somewhat weaker, allowing for more SN activity at lower redshifts and therefore somewhat stronger cores. However, as discussed in Sections~\ref{subsec:res_dm} and \ref{subsec:results_centraldm}, the differences induced by AGN feedback are only minimal for the \textbf{m10y} profiles at $z=0$.

For \textbf{m10z}, we obtain the most complex interplay between star formation, AGN activity and CRs. Here, the set-ups without CRs have more powerful initial star formation bursts (underlining that the initial intensity of the burst is not necessarily directly dictated by the galaxy formation physics configuration). Overall, \zF, \zC, and \zA \ all experience episodes of star formation for most of cosmic time, although for \zF, those are shifted to lower redshifts, whilst for the \zC \ simulation SN activity is more concentrated at higher redshifts. \zA \ experiences bursty star formation throughout cosmic time, leading to the highest cumulative injected SN energy and the most pronounced core. Likewise, the cumulative SN energy in the \zF \ and \zC \ set-ups also exceeds the $z=0$ binding energy and both of these result in cored profiles. Interestingly, however, star formation in the \zCA \ set-up is significantly suppressed, and powerful bursts are mostly restricted to high redshifts ($z>2$), so that the total injected SN energy remains well below the $z=0$ binding energy by a factor of three. This indicates that for the  \zCA \ set-up, the core formation may be driven by the AGN.

This analysis underlines that CRs and AGN primarily affect cusp-to-core transformation, and the diversity of dark matter profiles, by regulating star formation histories in our simulations. The only set-up where the AGN feedback is actively driving core formation is \zCA, where star formation is too heavily suppressed to drive core formation. In the cases of \textbf{m10q} and \textbf{m10y}, the AGN is not efficient at suppressing star formation, leading to similar or higher levels of SN activity compared to the equivalent no-AGN set-ups. However, in these scenarios, the AGN may still enhance core formation; we examine this possibility in the next section.

\subsubsection{AGN feedback} \label{subsubsec:AGN_driver}

\begin{figure*}
    \centering
    \includegraphics[width=\textwidth]{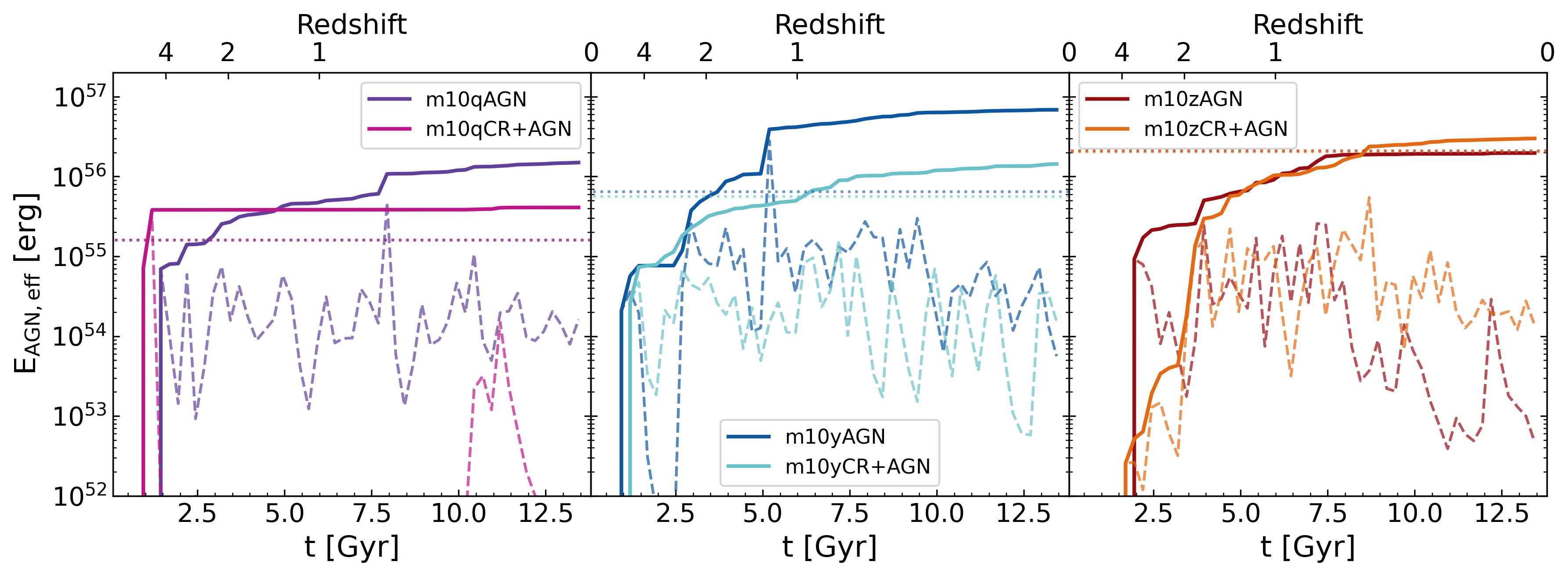}
    \caption{Instantaneous (dashed lines) and cumulative (solid lines) effective AGN energy injected as a function of cosmic time. The three panels display simulations based on our three sets of ICs, with colour-coding indicating the galaxy formation physics configurations as listed in the legends. The $z=0$ binding energy of the halos is indicated by colour-coded dotted lines. AGN feedback is injected through three channels: radiation, mechanical winds, and, for some, CRs. The overall ``effective'' energy is calculated based on reported couplings from the literature. Runs with CRs show more efficient AGN feedback regulation, with \zCA \ experiencing efficient bursts that induce a core despite low star formation activity. Without CRs, BHs grow more efficiently, leading to higher energy injection rates and increased core sizes for all set-ups.}
    \label{fig:agn_egy}
\end{figure*}

In the \textsc{fire-3} model, AGN feedback is injected in several channels including mechanical feedback, radiative feedback and CRs (see Section~\ref{subsubsec:methods_bhs} for details). Comparing this multi-channel AGN energy to the SN energy that is predominantly injected mechanically is not straightforward, especially since we are most concerned with feedback-induced gravitational potential perturbations. Therefore, we need to assess how the AGN energy will couple dynamically. Fully tracking and analysing coupling efficiencies of different AGN feedback modes is beyond the scope of this paper, so we make some simplifying assumptions based on the previous literature.

For the AGN, only a very small fraction of the injected energy is directly coupled in the form of mechanical winds, with only $\sim0.6$ per cent of the AGN luminosity injected as wind energy (see Section~\ref{subsubsec:methods_bhs}). 

For the radiative feedback, on the other hand, the full luminosity is injected into the simulation domain using the LEBRON method \citep[see][]{hopkins_radiative_2020} for radiation transport. This radiation transport then predicts the radiation pressure forces, i.e. momentum flux, from the absorbed photon luminosity. Overall only a small fraction of the luminosity will therefore be energetically coupled to the gas as a radiatively driven wind. The LEBRON method has been most thoroughly tested for stellar feedback. Whilst AGN radiation has a different spectrum, we can still draw some tentative conclusions from these studies for our estimates. For the stellar feedback, only $0.4 L/\mathrm{c}$ of the photon momentum couples to the gas in dwarfs, whilst the remainder escapes the galaxy, and the vast majority of this photon coupling occurs in the single-scattering regime \citep[see analysis in][]{hopkins_radiative_2020}. We here assume that a similar fraction of the luminosity would couple to the gas for radiative AGN feedback. Following \citep{king_powerful_2015}, a fraction of approximately 5 per cent of that availabile AGN luminosity is then coupled as mechanical energy in the form of a radiatively driven wind in the single scattering regime, yielding an overall mechanical energy efficiency of 2 per cent compared to the AGN luminosity for the radiative feedback channel.

For the AGN-driven CR feedback, 1 per cent of the accreted rest-mass energy, i.e. 10 per cent of the luminosity with the assumed radiative efficiency of 10 per cent, is injected as CRs. For CR transport, we employ the subgrid model from \citet{hopkins_simple_2023}, which interpolates between the limit in which CRs escape the galaxies with negligible losses and that in which CRs lose most of their energy catastrophically before escaping, using a formalism akin to the LEBRON method. For dwarf galaxies, we are generally far from the proton calorimetric limit where most energy is lost before CRs escape the dense gas, and we assume that (as with the radiation) 40 per cent of the CRs couple before escaping -- which likely provides an upper limit. This then yields an overall mechanical energy efficiency of 4 per cent compared to the AGN luminosity for the CR feedback channel.

Keeping in mind that these assumed efficiencies may be overestimating the actual coupled AGN energy, we show the `effective' injected AGN energy based on our assumed efficiencies for the different channels as a function of cosmic time in Fig.~\ref{fig:agn_egy}. Again the cumulative energy is shown as the solid lines and the instantaneous energy is shown as dashed lines binned over 250~Myr for clarity. The AGN energy is based on the accretion rate of the most massive BH in the main halo at $z=0$ and its most massive progenitor halo at higher redshifts. The three different panels represent our three different ICs and the colour coding of the lines indicates the different physics configurations as listed in the legends.

Overall, we note that AGN energy injection is more steady than SN energy injection, with the AGN in most set-ups being active continuously, except for \qCA \ where the AGN activity, just like the star formation, is quenched at early times and is then mostly quiescent at late times apart from a relatively weak late-time burst. In all cases, the injected AGN energy matches or exceeds the binding energy though we caution that due to the extremely uncertain mechanical coupling efficiencies this has to be taken with a grain of salt and the AGN energy that is actually coupled mechanically to the gas may be lower. Furthermore, we note that whilst the SN feedback is distributed throughout the galaxy the AGN energy injection is only focussed at the centre, with radiation and mechanical winds injected in a collimated fashion, and therefore AGN feedback is likely less efficient than SN feedback in affecting the ISM at a fixed energy injection rate. Indeed, past simulations have demonstrated that a significant fraction of the AGN wind energy escapes the ISM once the winds have opened up a central cavity \citep[e.g.][]{bourne_resolution_2015,koudmani_fast_2019,torrey_impact_2020,mercedes-feliz_local_2023}.

For \textbf{m10q}, in the CR-based run \qCA, the AGN does not have a significant impact on the central densities since the late-time activity, which is crucial for core formation, is mostly suppressed. Without CRs, for the \qA \ set-up, there is a sharp AGN burst at $t \sim 8$~Gyr which is also associated with pronounced central dark matter density fluctuations and followed by a steady decline in the central dark matter density suggesting that the AGN may be contributing to the formation of the core with this set-up.

For \textbf{m10y}, the \yA \ set-up has a strong AGN burst at $t\sim 5.5$~Gyr, which is again associated with strong central density fluctuations indicating gravitational perturbations as well as a drop in the star formation rate. With the CRs, the AGN energy injection is again significantly suppressed (though not completely shut down like in the \qCA \ set-up) and the relatively low levels of activity do not have a significant impact on core formation.

For \textbf{m10z}, contrary to the other two sets of ICs, the energy injected from the CR-based run, \zCA, exceeds the AGN energy injected with the equivalent no-CR set-up \zA. This is mostly driven by a significant burst in the \zCA \ set-up at $t\sim 9$~Gyr, which takes the overall injected AGN energy budget above the binding energy of the halo and is followed by strong density fluctuations. Most notably, even though central star formation is completely suppressed from $t\sim 10$~Gyr in the \zCA \ simulation, the AGN activity persists at relatively high levels, maintaining the core in the absence of SN-driven winds. This is driven by two factors in our AGN modelling. Firstly, at late times the ratio between BH mass and subgrid accretion disc mass in the \zA \ run is quite large ($M_\mathrm{BH}/M_\mathrm{d} \sim 10^{4}$), which leads to depletion timescales of the order of $\sim 1$~Gyr (also see Section~\ref{subsubsec:methods_bhs}), allowing for persistent AGN activity even after the inflows onto the BH -- accretion disc particle have subsided. Secondly, whilst the gas density is significantly decreased in the central region (and therefore the gas is not star-forming), the BH is still able to accrete at low rates from this supply since there is no explicit density criterion for the BH accretion. This likely represents an optimistic estimate of the BH accretion rate and we discuss the implications and caveats of these modelling choices in more detail in Section~\ref{subsubsec:disc_bhs}. For the \zA \ set-up the AGN has a significant burst at $t\sim12$~Gyr which again is associated with strong central density fluctuations suggesting that here the AGN may also enhance the formation of the core.

\begin{figure*}
    \centering
    \includegraphics[width=\textwidth]{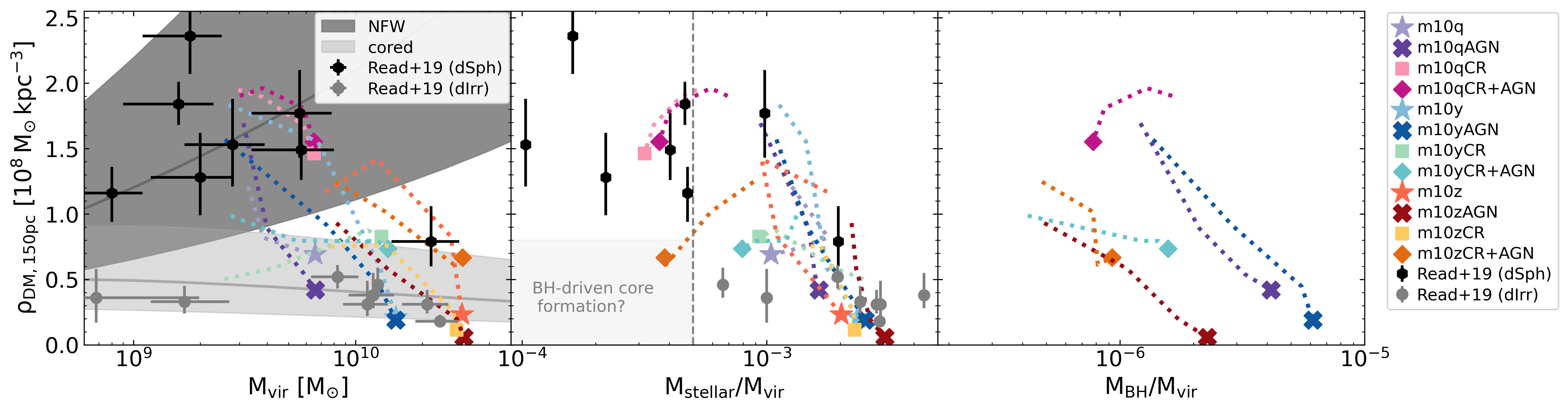}
    \caption{Central dark matter densities at $r=150$~pc as a function of virial mass (\textit{left panel}), stellar mass to virial mass ratio (\textit{middle panel}), and BH mass to virial mass ratio (\textit{right panel}). Simulated $z=0$ data points are highlighted by marker symbols corresponding to the respective galaxy formation physics configurations as indicated by the legend. The binned redshift evolution from $z=3$ to $z=0$ is represented by dotted lines. The left and middle panels include various observed central dark matter densities from \citet{read_dark_2019} for comparison. In the middle panel, a dashed grey line marks the stellar mass to virial mass ratio where previous studies suggest stellar-feedback-driven core formation may become inefficient. Our simulations broadly agree with this theory, except for \zCA, which forms a core despite a decreasing stellar mass to virial mass ratio, though has an increasing BH mass to virial mass ratio, indicating AGN-driven core formation.}
    \label{fig:dm_central_dens}
\end{figure*}

Overall, we note for most of our AGN set-ups, including \qA, \yA, \yCA, \zA, and \zCA, the equivalent no-AGN set-up already has a SN-induced core. For all of these cases, apart from \zCA, the addition of an AGN further decreases the central densities enhancing the existing core. However, it is difficult to disentangle whether this additional decrease is driven directly by AGN energy injection or indirectly by suppressing high-redshift star formation and `delaying' SN activity to later times favouring core formation. Indeed, it is likely that both of these factors contribute to the enhancement of cores.

The \zCA \ set-up is our only simulation where the AGN is efficient in globally suppressing star formation compared to its no-AGN counterpart (also see Fig.~\ref{fig:stellar_prop}). Here, there is no SN activity for the last $\sim3.5$~Gyr from $z\sim 0.5$ and only weak SN activity between $z=2$ and $z=0.5$, yet the core is maintained. Indeed, the central density significantly decreases from $z=2$ to $z=1$ and is then maintained at low levels despite the late-time merger at $z \sim 0.5$, providing strong evidence for AGN-driven core formation.

Overall, we conclude that AGN feedback can induce significant scatter in dark matter density profiles,  predominantly indirectly by regulating star formation histories but also in some cases directly by driving powerful potential fluctuations.

\subsection{Cosmic evolution of central dark matter densities in observational context} \label{subsec:results_centraldm}

We further analyse the origin of the diversity of our simulated dark matter profiles by examining the central densities in the context of observations. Fig.~\ref{fig:dm_central_dens} shows the central dark matter densities at $r=150$~pc as a function of virial mass (left panel), stellar mass to virial mass ratio (middle panel) and BH mass to virial mass ratio (right panel). The dotted lines indicate the (binned) redshift evolution of our simulated dwarfs from $z=3$ to $z=0$, again binned over 500~Myr. For clarity we only highlight the $z=0$ data point with the marker symbol corresponding to our different physics configurations, as indicated by the legend. In the left panel we also show the expected central densities for an NFW profile and a cored profile following the \textsc{coreNFW} profile from \citet{read_dark_2016} as grey lines with the $2\sigma$ concentration scatter indicated by dark-grey and light-grey shaded regions, respectively. Furthermore, we show the observed central densities of 16 observed nearby dwarf galaxies, including classical dwarf spheroidals and gas-rich dwarf irregulars from \citet{read_dark_2019}.

The first panel demonstrates that our simulations are generally in good agreement with the observed central densities of nearby dwarfs, keeping in mind that the sample sizes are limited in both cases (16 observed dwarfs versus 12 simulated dwarfs). Two of the \textbf{m10z} simulations, \zC \ and \zA, lead to very strong cores, with the central densities being more severely suppressed than in the observations. Our simulations do not reproduce the observed strong cusps with central densities above the NFW mean relation. However, we note that the majority of our dwarfs have late star formation histories, with only the fiducial and CR set-ups for \textbf{m10q} and CR set-ups for \textbf{m10y} resulting in an early truncation of star formation, as defined by \citet{read_dark_2019}, with no (significant) activity for the past 6~Gyr. These dwarfs with early truncation are associated with cuspier profiles hinting that we would require more early-forming set-ups to reproduce these strong cusps.

The middle panel shows the simulated and observed central densities as a function of stellar mass to virial mass ratio $M_\mathrm{stellar}/M_\mathrm{vir}$. As discussed in \citet{read_dark_2019}, the observed dwarfs show a clear trend with higher $M_\mathrm{stellar}/M_\mathrm{vir}$ being associated with lower central dark matter densities. Moreover, in the observations, as well as in previous theoretical studies, it is found that no stellar-feedback-driven cores form below a critical ratio of $\left( M_\mathrm{stellar}/M_\mathrm{vir} \right)_\mathrm{crit}= 5 \times 10^{-4}$ since the integrated SN energy is insufficient to drive the required potential fluctuations \citep[see discussion in][]{di_cintio_dependence_2014}. We also find a strong correlation between $M_\mathrm{stellar}/M_\mathrm{vir}$ and central density suppression for our simulated dwarfs. Indeed, the correlation is significantly tighter than for the observed dwarfs likely due to increased scatter induced by measurement uncertainties in the observations. Looking at the redshift evolution tracks of our simulated dwarfs, we find that most systems also follow these trends in a temporal sense with two notable exceptions. The CR-based runs with the \textbf{m10q} ICs show a slight decrease as the $M_\mathrm{stellar}/M_\mathrm{vir}$ decreases. However, this is mostly a by-product of these runs being quenched at very high redshifts so that the central dark matter density evolution is dominated by the general assembly history of the halo with the virial mass increasing and the halo concentration decreasing (whilst the stellar mass remains constant) at late times. The second exception to the $M_\mathrm{stellar}/M_\mathrm{vir}$ trend, however, is more notable: for the \zCA \ set-up the central densities as a function of $M_\mathrm{stellar}/M_\mathrm{vir}$ are steadily decreasing so that a core forms despite the $z=0$ $M_\mathrm{stellar}/M_\mathrm{vir}$ ratio being below the critical ratio identified by \citet{di_cintio_dependence_2014} and \citet{read_dark_2019}. This provides further evidence that this run could be experiencing AGN-driven core formation.

We further examine this possibility in the third panel where we plot the simulated central densities (as well as their binned redshift evolution from $z=3$) for the AGN-based simulation set-ups as a function of BH mass to virial mass ratio $M_\mathrm{BH}/M_\mathrm{vir}$. For the runs without CRs, \qA, \yA, and \zA, the central densities again steadily decrease as a function of $M_\mathrm{BH}/M_\mathrm{vir}$. However, we caution that this does not necessarily indicate that the BHs are contributing to core formation due to the strong correlation between $M_\mathrm{BH}$ and $M_\mathrm{stellar}$ (see Fig.~\ref{fig:scal_rel} for scaling relations of our simulated dwarfs). The \qCA \ set-up is the only simulation where the central density ratio decreases as $M_\mathrm{BH}/M_\mathrm{vir}$ decreases, as with the stellar mass to halo mass ratio, and this mainly reflects that the central density is unaffected by baryonic processes at late times. The \yCA \ set-up has a very shallow gradient again reflective of the relatively quiescent late-time evolution. Finally, the \zCA \ set-up exhibits a significant decrease in central densities as the $M_\mathrm{BH}/M_\mathrm{vir}$ increases (opposite trend from $M_\mathrm{stellar}/M_\mathrm{vir}$) providing further support for the BH-driven core formation scenario.

\subsection{The diversity of rotation curves} \label{subsec:res_diversity}

As discussed in Section~\ref{sec:intro}, one of the most prominent remaining dwarf galaxy `problems' pertains to the observed diversity of dwarf galaxy rotation curves. From our analysis in the previous sections, we have found that AGN and CRs may significantly increase the scatter in dark matter density profiles and hence may contribute to resolving this on-going controversy. Ultimately, to make statistical predictions for dwarf galaxy rotation curve shapes, we would need a much larger simulation sample spanning a larger range of environments and halo masses. Nevertheless, we can use our simulation suite to examine the broad trends for rotation curve shapes and assess the potential for more sophisticated galaxy formation models including the impact of AGN and CRs at the low-mass end to resolve the long-standing diversity of rotation curves problem.

\begin{figure*}
    \centering
    \includegraphics[width=\textwidth]{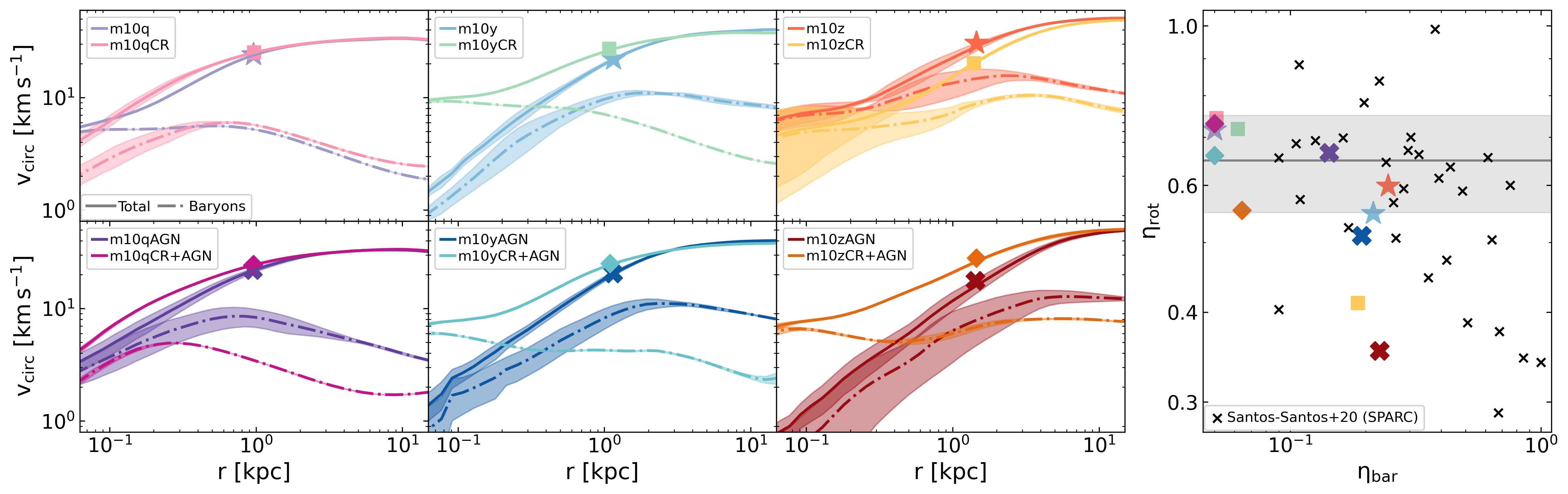}
    \caption{Circular velocity profiles at $z=0$ as a function of radius including all matter (solid lines) and baryons only (dashed-dotted lines). The profiles are averaged over $\Delta T=500$~Myr with the shaded regions indicating the $1\sigma$ scatter. The first three columns display simulations based on our three sets of ICs with the no-AGN runs (top row) and AGN runs (bottom row) plotted separately for clarity. The markers highlight the circular velocity at the fiducial radius which is used as a reference point in observational studies of rotation curves. The fourth column translates these velocity curves to the rotation curve shape -- baryon dominance parameter space, also showing observed dwarf galaxies in a similar mass range from the SPARC sample as compiled in \citet{santos-santos_baryonic_2020}.  Cosmic environment, CRs and BH feedback can all influence the rotation curve shape parameter, with our simulations broadly reproducing the observed trends with baryonic dominance.}
    \label{fig:dm_vcirc}
\end{figure*}

In Fig.~\ref{fig:dm_vcirc}, we present circular velocity curves for all our dwarf simulations. As for the dark matter profiles, we average the velocity curves over $\Delta T=500$~Myr and indicate the $1\sigma$ scatter with the shaded regions. Following the methodology presented in \citet{santos-santos_baryonic_2020}, we plot circular velocity profiles for the total matter content of our dwarfs (dark matter, gas, stars, and BHs) as solid lines and for baryons (gas, stars, and BHs) as dotted-dashed lines.

The total circular velocity curves for each set of ICs converge between 1 to 10 kpc, with the extent of the matter deficits in the central regions reflecting the core sizes from Fig.~\ref{fig:dm_prof}. For the baryonic circular velocity curves, however, there are significant differences. We also examine the gas and stellar distributions (not shown here) and find that both of these are significantly modified for galaxies with pronounced cores, which tend to have more extended stellar distributions (also see Figure~\ref{fig:proj_suite}). For the cored dwarfs this then leads to slowly rising rotation curves for both dark matter and stars. Dwarfs that are quiescent at $z=0$, and generally have cuspier profiles, are more gas poor (especially at the centre), leading to overall lower baryonic circular velocities in the central region.

Following \citet{santos-santos_baryonic_2020}, we quantify the cusp versus core nature of the profiles by calculating characteristic velocity ratios, comparing the inner `fiducial' rotation velocity $V_\mathrm{fid}$ with the asymptotic flat rotation velocity $V_\mathrm{max}$. The fiducial rotation velocity is measured at the fiducial inner radius defined as:
\begin{equation}
    r_\mathrm{fid} = \left(\frac{V_\mathrm{max}}{35 \ \mathrm{km\, s^{-1}}} \right) \ \mathrm{kpc}.
\end{equation}
The rotation curve shape parameter $\eta_\mathrm{rot}$ can then be defined as 
\begin{equation}
    \eta_\mathrm{rot} = \frac{V_\mathrm{fid}}{V_\mathrm{max}} = \frac{V(r_\mathrm{fid})}{V_\mathrm{max}}.
\end{equation}
Rapidly rising rotation curves have $\eta_\mathrm{rot} \lesssim 1$, with the NFW profile yielding a rotation curve shape parameter of $\eta_\mathrm{rot} \sim 0.65$. Cored profiles have $\eta_\mathrm{rot} \ll 1$; see \citet{santos-santos_baryonic_2020} for a detailed discussion of this parameter.

As in \citet{santos-santos_baryonic_2020}, we also inspect a second velocity ratio -- the baryonic importance parameter, which is defined as
\begin{equation}
    \eta_\mathrm{bar} = \left( \frac{V_\mathrm{b,fid}}{V_\mathrm{fid}}\right)^{2} = \left( \frac{V_\mathrm{b}(r_\mathrm{fid})}{V (r_\mathrm{fid})}\right)^{2},
\end{equation}
where $V_\mathrm{b,fid}$ is the contribution from baryons (gas, stars, BHs) to the circular velocity curve. Under the assumption of spherical symmetry, $\eta_\mathrm{bar}$ represents the mass fraction of baryons within the fiducial radius.

The two velocity ratios $\eta_\mathrm{rot}$ and $\eta_\mathrm{bar}$ can be used to plot the `cuspiness' of the rotation curves against the contribution from baryons in the inner region of the galaxy. We plot this distribution in the rightmost panel in Fig.~\ref{fig:dm_vcirc}. For comparison, we also show observed data as collated by \citet{santos-santos_baryonic_2020}. We restrict the observed galaxy sample to the classical dwarf regime with $V_\mathrm{max} \leq 60 \ \mathrm{km \, s^{-1}}$ to compare this more easily and directly with our simulated dwarfs. The observational data stems mainly from the SPARC sample and, in all cases, the rotation curves were inferred from high-resolution HI and/or H$\alpha$ velocity fields. Given that all of the observational rotation curves are inferred from the baryonic contributions, the dwarfs with extremely low baryonic contributions are missing from the observed samples. Similarly to \citet{santos-santos_baryonic_2020} we therefore show our low-$\eta_\mathrm{bar}$ simulated dwarfs at $\eta_\mathrm{bar} = 0.05$ for clarity. 

Focussing first on the observations, we find that, as discussed in \citet{santos-santos_baryonic_2020}, the baryonic importance parameter appears to be a \textit{poor predictor} of rotation curve shapes with a significant amount of scatter in the data. Specifically, this plot highlights the appearance of profiles ranging from cusps to weak cores at low baryonic dominance and cusps to strong cores at high baryonic dominance in the observations. This large amount of scatter may naively seem at odds with the theory that baryons are driving cusp to core transformations and has been extremely difficult to reproduce with simulations. Despite the high scatter, there are nevertheless some interesting observational trends where strong cores in dwarfs tend to be associated with higher baryonic importance and strong cusps tend to be associated with lower baryonic importance. However, both of these aspects seem counter-intuitive if cores are associated with strong outflows and cusps with baryons cooling at the centre. 

Our simulations cover almost all observed scenarios: cusps and weak cores at low baryonic dominance (e.g., \qC \ and \zCA) and cusps and strong cores at high baryonic dominance (e.g., \yF \ and \zA). The only missing scenario are the `ultra cuspy' dwarfs at both low and high baryonic dominance, which might be influenced by environmental factors, suggesting our sample might be too small to cover the required range of assembly histories. To reproduce these extremely cuspy dwarfs with CDM-based simulations, we would need strong cosmic inflows so that feedback becomes inefficient leading to overcooling \citep[see e.g.][]{smith_cosmological_2019}. Another possible pathway for inferring ultra-cuspy dwarfs from gas rotation curves constitutes strong, transient perturbation to the gas, usually resulting in an offset between the gas, stars and dark matter. However, in the fiducial \textsc{fire-3} model (as well as in the SPARC sample), these transient ultra-cuspy dwarfs are generally associated with more massive systems ($M_\mathrm{halo} \gtrsim 10^{11} \ \Msun$ and $V_\mathrm{max} \gtrsim 50 \ \mathrm{km \, s^{-1}}$), see \citet{sands_confronting_2024} for details.

Nevertheless, even with our small sample, we obtain a range of profiles at fixed baryonic importance ranging from cusps to cores, a feat that has been historically difficult to reproduce in simulations \citep[e.g.][]{oman_unexpected_2015,garrison-kimmel_local_2019,santos-santos_baryonic_2020}. We are able to reproduce these trends due to the interplay of the different feedback processes resulting in varying overall feedback efficiencies for the same baryon content, either enhancing or reducing the overall feedback impact compared to simulations that do not include CRs or BHs. Cusps are generally associated with highly concentrated stellar distributions in gas-poor dwarfs, whilst cores are created by galactic fountain activity in gas-rich dwarfs that produces a core through cyclic feedback events but does not expel the baryons. In this manner, we are able to match both the large observed scatter and the weak observed trends with baryonic dominance.

\section{Discussion} \label{sec:discussion}

\subsection{Comparison with previous theoretical work}

We have investigated the impact of BH feedback and CRs on the cusp versus core problem in dwarf galaxies with a special focus on whether including additional baryonic feedback channels may help explain the observed diversity of rotation curves in dwarf galaxies.

Several groups have investigated the role of stellar feedback in driving cusp to core transformations in galaxy formation simulations with mixed results \citep[e.g.][]{governato_bulgeless_2010,parry_baryons_2012,pontzen_how_2012,hopkins_galaxies_2014,hopkins_fire-2_2018,kimm_towards_2015,emerick_stellar_2018,smith_cosmological_2019,gutcke_lyra_2021}. Most simulations tend to either predominantly produce cusps or predominantly produce cores which appears to be closely related to the star formation threshold employed and, more crucially, whether the cold, dense ISM phase is resolved in the simulations \citep[see discussion in][]{jahn_real_2023}. Nevertheless, for a given ISM model it has proven very challenging for simulations to reproduce the observed diversity of dwarf rotation curves \citep[e.g.][]{oman_unexpected_2015,garrison-kimmel_local_2019,santos-santos_baryonic_2020}. The past decade has unveiled growing observational samples of dwarf galaxies with AGN. These observations raise the question of whether active BHs may increase the diversity in star formation histories as well as directly impact central dark matter distributions and thereby increase the diversity of dwarf rotation curves in simulations.

Previous studies found systematic yet small-scale suppressions of central dark matter densities by AGN feedback in dwarfs \citep{koudmani_two_2022,arjona-galvez_role_2024}, yet in both cases, the simulations did not explicitly resolve the multiphase ISM. Cusp-to-core transformations with AGN feedback were also explored by \citet{arora_dark_2024}. However, they employed the fiducial Bondi accretion model for BH growth, which suppresses the growth of low-mass BHs such that AGN activity in dwarfs is insignificant in these simulations and does not affect dark matter distributions.

In our study, we investigate, for the first time, the impact of AGN feedback on dark matter profiles in dwarfs with a resolved multi-phase ISM model and a BH growth scheme that does not suppress AGN activity in dwarf galaxies. Several works exploring BH physics within the \textsc{fire} model have found that CRs appear to be a key ingredient for effective AGN feedback \citep[e.g.][]{su_which_2021,wellons_exploring_2023,byrne_effects_2023}, so we also investigate the impact of CR feedback on the interplay between baryons and dark matter in dwarfs. We find that efficient AGN feedback can actively drive core formation and, together with CRs, significantly enhances the diversity of dwarf rotation curves by leading to more varied star formation histories. Indeed, in many cases our runs including CRs (with or without AGN) lead to cuspier profiles due to suppressed star formation activity. This is in contrast with recent findings from \citet{martin-alvarez_pandora_2023}, who find that CRs are prone to enhance core formation. We note that for one of our set-ups (compare \zF \ to \zC), we also observe this behaviour, and indeed it is plausible that CRs could act to either suppress or enhance cores depending on their interaction with the ISM and the assembly history of the given dwarf galaxy. Our most important finding is that BHs and CRs as additional baryonic processes lead to varying overall feedback efficiencies that can lead to both cuspier or more cored profiles for a given assembly history. 

We also demonstrate the promising potential for these additional baryonic processes to resolve the diversity of dwarf galaxy rotation curves problem: our simulations exhibit a wide variety of profiles, including cuspy profiles at low baryonic dominance and strong cores at high baryonic dominance, thereby reproducing the puzzling relationship between rotation curve shapes and the gravitational importance of baryons from the observations. The former is associated with highly concentrated stellar distributions in gas-poor dwarfs, whilst the latter is a signature of galactic fountain activity in gas-rich dwarfs that produces a core through cyclic feedback events but does not expel the baryons.

One remaining discrepancy with the observations lies in the `extremely' cuspy dwarfs which are even denser than NFW expectations. Within our relatively modest sample size of simulations we are unable to reproduce these. Apart from the obvious need to explore more halo masses and environments, as these objects may be linked to strong cosmic inflows which lead to overcooling \citep[see e.g.][]{smith_cosmological_2019}, in future work, it would also be important to explore observational uncertainties in rotation curve measurements as well as theoretical uncertainties in the modelling of BHs and CRs, and we discuss both of these aspects in the following sections.

\subsection{Observational uncertainties}

In observations, the dark matter distribution of dwarf galaxies may be inferred from the rotation curves of gas or stars, yet it has been shown that in practice these rotation curves may be highly unreliable tracers of the `true' underlying circular velocity distribution. In particular, various dynamical perturbations may lead to large discrepancies of 50 per cent or more \citep[see][for some recent studies]{pineda_rotation_2017,roper_diversity_2023,downing_many_2023,sands_confronting_2024}. Processes that are associated with inducing AGN activity, including mergers and strong cosmic gas inflows, lead to inaccurate estimates of the matter distributions in dwarfs. Interestingly, all of these studies find that observational uncertainties are more likely to lead to an underestimation of the true circular velocity, i.e. overestimation of the occurrence of cored profiles in observations, whilst the inverse error is less common though may still occur in the central few kiloparsecs due to dynamical phenomena, especially for more massive dwarfs \citep[see discussion in][]{sands_confronting_2024}.

Apart from the observational uncertainties in inferring dark matter mass distributions, it also remains extremely challenging to determine the drivers behind core formation from observations. In particular, there are no observational constraints on the potential role of AGN in cusp-to-core transformations. This is mainly due to the number of observed AGN in nearby dwarf galaxies being very limited. To further complicate matters, AGN are associated with disturbed gas rotation curves -- see e.g. \citet{manzano-king_active_2020}, who find a strong association between AGN activity and disturbed gas kinematics in observed dwarf rotation curves, making it even more difficult to establish a link between AGN activity and central dark matter densities. Current and upcoming IFU surveys may partly alleviate these issues by disentangling rotational and outflow components, e.g. using MaNGA \citep[see e.g.][]{rodriguez_morales_manga_2025} or high-resolution NIRSpec-IFU observations with JWST \citep[e.g.][]{bohn_big_2023}. From our simulations, we predict that if cored dwarf galaxies were found below the critical mass ratio of $\left(\frac{M_\mathrm{stellar}}{M_\mathrm{vir}}\right)_\mathrm{crit} = 5 \times 10^{4}$ this would be a strong indicator of AGN-driven core formation in dwarfs. 

\subsection{Theoretical uncertainties}
The diversity of models presented in this work (with or without CRs and with or without AGN) are meant to represent different outcomes of `the same underlying physics' under different conditions. In particular, the impact of AGN feedback is expected (and observed) to be highly variable in the dwarf regime. The AGN activity levels in dwarfs are highly dependent on the seeding model assumed (with some dwarfs possibly not hosting any BHs at all) and whether the BH is off-centre for the majority of the dwarf's history, i.e. the no-AGN model here may represent the same underlying physics as the AGN model but in a situation where no BH seeded or only an extremely weakly accreting off-centre BH is present. Hence just seeding and BH dynamics could drive substantial variability in addition to other effects such as varying feedback efficiencies depending on BH spin evolution. Similarly, the CR subgrid model we employ here represents a case where losses within the galaxy are minimal and hence the impact of CRs in our dwarfs likely represents an upper limit. With full CR transport the outcome will likely lie somewhere in-between the no-CR and CR simulations presented in this work. It would require significantly higher resolution simulations (and correspondingly more detailed BH modelling) and explicit CR transport to depict the full variability in dark matter distributions introduced by baryonic physics within \textit{one} galaxy formation model. Our simulations provide strong motivation for exploring such detailed models in future work. Below we discuss the caveats of our CR and BH modelling in more detail.

\subsubsection{CR modelling}
Whilst the focus of our paper lies on the interplay between AGN feedback and dark matter distributions in dwarfs, we also assess the impact of CRs on the cusp versus core problem since CR injection has been identified as an important AGN feedback channel. For this, we take advantage of the CR subgrid model by \citet{hopkins_simple_2023}, which allows for the inclusion of the impact of CRs on galaxy formation without the significant computational overhead associated with full CR transport.

As discussed in previous sections, for dwarf galaxies, this subgrid model presents a good approximation since, in this mass regime, we are far away from the calorimetric limit for protons. Consequently, we can safely assume that losses within the galaxy only play a secondary role. Nevertheless, with this subgrid model, we may still be somewhat overestimating the effects of CR feedback at late cosmic times, as we are not fully capturing the inhomogeneity of CRs in the CGM, which may modify thermal instabilities \citep{butsky_impact_2020}. In particular, we find that this implementation of CR feedback suppresses star formation so strongly that the addition of CRs generally leads to cuspier profiles. Indeed \citet{hopkins_simple_2023} also find in their validation simulations with a $10^{11} \ \Msun$ halo that the subgrid CR model leads to suppressed star formation and a somewhat weaker core compared to an equivalent simulation with full CR transport, though these differences are small ($\sim 1$~per cent) and both CR models lead to a weaker core compared to their no-CR set-up. Our results should therefore be regarded as an upper limit for the impact of CR feedback and, whilst exploring alternative CR implementations is beyond the scope of this paper, our work provides strong motivation for investigating cusp-to-core transformations with AGN and full CR transport in future work.

\subsubsection{BH modelling} \label{subsubsec:disc_bhs}
The modelling of AGN feedback in cosmological simulations remains subject to significant theoretical uncertainties due to the vast dynamic range of relevant scales spanning (at least) 14 orders of magnitude from the event horizon ($\sim 10^{-6}$~pc for Sgr A*) to the cosmic web ($\sim 10^{8}$~pc). Hence BH growth, feedback, and dynamics cannot be modelled ab-initio and need to be included as `subgrid' processes in cosmological simulations. 

Here, we model the BH growth based on the torque-based accretion model \citep{hopkins_analytic_2011}, which allows for efficient BH growth in the dwarf regime because it does not suffer from the strong BH mass dependence inherent to the widely used Bondi model \citep[also see][]{angles-alcazar_black_2013,angles-alcazar_torque-limited_2015,koudmani_two_2022,wellons_exploring_2023}. This leads to relatively high BH masses compared to the extrapolated scaling relations -- although we emphasise again that this comparison should be interpreted with caution because we would expect a flattening of the scaling relations in this mass regime \citep{greene_intermediate-mass_2020}. Nevertheless, from an accretion perspective, our simulations may be exploring an upper limit on the impact of BHs in the dwarf regime.

The inflow rates provided by the torque-based model are then coupled to a gas reservoir representing the accretion disc, with the depletion time of this reservoir set by the thin $\alpha-$disc model. However, there are a few caveats to this approach. Firstly, the $\alpha-$disc is generally only applicable in the radiatively efficient accretion regime. Secondly, this approach only tracks the mass flow rates through the disc and does not follow angular momentum flow rates. This in turn means that the BH spin evolution cannot be tracked self-consistently; therefore we have to assume constant efficiencies for the disc and jet luminosities, which otherwise could be directly inferred from the disc and BH properties \citep[see e.g.][]{beckmann_dense_2019,sala_non-isotropic_2021,talbot_blandford-znajek_2021,talbot_blandford-znajek_2022,talbot_simulations_2024,husko_spin-driven_2022, husko_winds_2024,koudmani_unified_2024}.

For the AGN feedback, we only explore one set of parameters for the three AGN channels (mechanical winds, radiation and CRs) that allows us to match the main observed characteristics of nearby dwarf galaxies; see Fig.~\ref{fig:stellar_prop}. However, as discussed in \citet{wellons_exploring_2023}, there are several other `plausible' AGN feedback set-ups within the \textsc{fire} model, leaving a large parameter space to be explored. In particular, we assume collimated injection for winds and radiation; however, the cusp versus core problem may be sensitive to alternative injection geometries, as discussed in \citet{zhang_bursty_2024} for the case of stellar feedback. This may allow us to also have a direct impact on dwarf core formation at lower halo masses, as in the \zCA \ set-up, without compromising in terms of producing realistic stellar properties.

The final caveat in our model is that we do not have sufficient resolution to self-consistently model BH dynamical friction. Hence we employ a drift force to model the unresolved dynamical friction, which generally ensures that the BHs remain close to the centre of the halo. However, in dwarf galaxies, off-centre BHs may actually be physical \citep[see e.g.][]{pfister_erratic_2019,bellovary_origins_2021,ma_seeds_2021} and in particular for massive BH binaries, the BHs orbiting around the centre could be an important contribution to core formation via core scouring. This has mainly been investigated in the context of massive elliptical galaxies \citep[see e.g.][]{rantala_formation_2018,rantala_simultaneous_2019}, however, these mechanisms may also translate to the dwarf regime and, in turn, the central dark matter densities of dwarfs are expected to affect binary shrinking times \citep[see][]{tamfal_formation_2018}. To also account for the dynamical effects of massive BHs on core formation in future work, it will be important to include the unresolved effects of dynamical friction as accurately as possible following dynamical friction estimators based on high-resolution N-body simulations \citep[e.g.][]{ma_new_2023,genina_calibrated_2024,partmann_difficult_2024}.

\section{Conclusions} \label{sec:conclusions}

In this work, we have investigated the impact of feedback from massive BHs on cusp-to-core transformations and the associated `diversity of rotation curves problem' in dwarf galaxies with a new suite of high-resolution cosmological zoom-in simulations based on the \textsc{fire-3} galaxy formation model. Our suite includes three different dwarf haloes spanning a range of masses and environments within the classical dwarf regime of $8 \times 10^{9} \ \Msun < M_\mathrm{halo} < 4 \times 10^{10} \ \Msun$. For each of these haloes, we investigate simulations with and without AGN as well as with and without CRs yielding four physics variations for each of the three haloes. We note that these different set-ups may represent different outcomes of the same underlying galaxy formation physics under different conditions, with the efficiency of both AGN and CRs expected to be highly variable in dwarf galaxies. Our main conclusions are as follows:
\begin{enumerate}
    \item AGN may drive core formation directly as an additional source of feedback dynamically heating the central dark matter distribution; see the \zCA \ set-up.
    \item AGN may also enhance core formation indirectly by suppressing high-redshift star formation and shifting SN activity to later times, which favours core formation at $z=0$; see e.g. the \qA \ and \zA \ set-ups.
    \item CRs may also indirectly affect cusp-to-core transformations by regulating star formation histories -- for most of our simulations, this leads to suppressed star formation and therefore suppressed core formation; see e.g. the \qC \ and \yC \ set-ups.
    \item Our simulation suite is in good agreement with observed dark matter central density distributions, broadly following the trends from the nearby dwarfs in \citet{read_dark_2019}. In particular, cores are generally more pronounced for higher stellar-to-halo mass ratios, $M_\mathrm{stellar} / M_\mathrm{vir}$. The only exception to this trend comes from the AGN-driven core in the \zCA \ set-up, which has a strongly suppressed $M_\mathrm{stellar} / M_\mathrm{vir}$ ratio and a cored profile. This parameter space is a key target for observational searches for cores induced by AGN feedback.
    \item BH feedback and CRs create a variety of cusps and cores in circular velocity profiles, with correlations between rotation curve shapes and baryonic influence that align with observations, potentially helping to resolve the diversity of rotation curves problem in dwarf galaxies.
\end{enumerate}

Overall, our findings suggest that AGN in dwarf galaxies can significantly impact central dark matter densities. BHs influence dwarf galaxy rotation curves through two key mechanisms: directly driving core formation via AGN feedback and indirectly regulating core formation by altering star formation histories and SN activity. CRs further contribute to this regulation by enhancing AGN feedback and suppressing star formation, thereby influencing core formation. The combined effects of AGN and CRs lead to a spectrum of central dark matter densities, reflecting the varying levels of AGN activity, star formation, CR influence, and their non-linear interaction. Due to the bursty star formation and accretion histories in dwarfs, AGN feedback is likely to occur in a highly stochastic manner; this may explain the wide range of dark matter central densities inferred from dwarf observations at fixed halo mass. Linking surveys of dwarf rotation curves with searches for AGN signatures will be crucial for future observations to elucidate the impact of BHs on dark matter profiles and to determine whether the diversity of rotation curve problem may be explained by baryonic processes.

\section*{Acknowledgements}
The authors would like to thank the anonymous referee whose comments improved the quality of this paper. The authors are grateful for helpful discussions with Vasily Belokurov, Martin Bourne, Jenny Greene, Vid Irsic and Sergio Martin-Alvarez. The authors would also like to thank Isabel Santos-Santos for providing the observational rotation curve data. The simulations presented in this work were run on the Flatiron Institute's research computing facilities (the Iron compute cluster), supported by the Simons Foundation.  SK has been supported by a Flatiron Research Fellowship, a Junior Research Fellowship from St Catharine's College, Cambridge and a Research Fellowship from the Royal Commission for the Exhibition of 1851. The Flatiron Institute is supported by the Simons Foundation. DAA acknowledges support by NSF grant AST-2108944, NASA grant ATP23-0156, STScI grants JWST-GO-01712.009-A and JWST-AR-04357.001-A, Simons Foundation Award CCA-1018464, and Cottrell Scholar Award CS-CSA-2023-028 by the Research Corporation for Science Advancement. Support for ISS was provided by NSF Collaborative Research Grant 2108318. SW received support from the NASA RIA grant 80NSSC24K0838.

\section*{Data Availability}
The data underlying this article will be shared on reasonable request
to the corresponding author.


\bibliographystyle{mnras}
\bibliography{CuspVsCorePaper} 




\appendix


\bsp	
\label{lastpage}
\end{document}